\documentclass[Journal,a4paper]{IEEEtran}

\usepackage[T1]{fontenc}
\usepackage{graphicx}
\usepackage{titlesec}
\usepackage{setspace}

\titlespacing\section{0pt}{7pt plus 1pt minus 1pt}{3pt plus 1pt minus 1pt}
\titlespacing\subsection{0pt}{6pt plus 1pt minus 1pt}{2pt plus 1pt minus 1pt}
\usepackage{amsthm,amsmath,amssymb,bm}
\usepackage{esint}
\usepackage{mathrsfs}
\usepackage{amsfonts}

\newtheorem{remark}{Remark}

\usepackage[caption=false]{subfig}
\usepackage{algorithm}
\usepackage[noend]{algpseudocode}
\usepackage{graphics}
\usepackage{epsfig}
\usepackage{cite}
\usepackage{color}
\usepackage{booktabs}
\usepackage{verbatim}
\usepackage{stfloats}
\usepackage{cases}
\usepackage{multirow}
\usepackage[inline]{enumitem}
\usepackage{ragged2e}
\usepackage{epstopdf}
\usepackage[colorlinks, urlcolor=blue, linkcolor=blue, anchorcolor=blue, citecolor=blue]{hyperref}

\DeclareSubrefFormat{myparens}{#1(#2)}
\newcommand{\transpose}{\mathsf{T}}
\newcommand{\hermconj}{\mathsf{H}}
\newcommand{\E}{\mathsf{E}}
\newcommand{\trace}{\mathtt{tr}}

\begin{document}
	
\title{Rethinking Mutual Coupling in Movable Antenna MIMO Systems: Modeling and Optimization}
\author{
    \IEEEauthorblockN{Tianyi~Liao,~\IEEEmembership{Graduate~Student~Member,~IEEE}, Wei~Guo,~\IEEEmembership{Member,~IEEE}, \\ Jun~Qian,~\IEEEmembership{Member,~IEEE}, Zixin~Wang,~\IEEEmembership{Member,~IEEE}, Shenghui~Song,~\IEEEmembership{Senior~Member,~IEEE}, \\ Jun~Zhang,~\IEEEmembership{Fellow, IEEE}, and Khaled~B.~Letaief,~\IEEEmembership{Fellow, IEEE}}
    \thanks{
    An earlier version of this paper was accepted by IEEE International Conference on Communications (ICC), Glasgow, Scotland, UK, May 2026~\cite{liao2026rethinking}.
    
    The authors are with the Department of Electronic and Computer Engineering, The Hong Kong University of Science and Technology (HKUST), Kowloon, Hong Kong. Emails: ty.liao@connect.ust.hk, \{eeweiguo, eejunqian, eewangzx, eeshsong, eejzhang, eekhaled\}@ust.hk.}}
\maketitle
\IEEEaftertitletext{\vspace{-1\baselineskip}}

\setstretch{0.96}

\begin{abstract}
    Movable antennas (MAs) have attracted growing interest for their ability to improve channel conditions via adaptive antenna movement. Nevertheless, such movement inevitably introduces mutual coupling (MC), whose impact has been largely overlooked in existing MA literature. In this paper, we show that MC is not merely an unavoidable electromagnetic effect, but also a new source of capacity gains in MA-enabled multiple-input multiple-output (MIMO) systems. To leverage MC effects, we develop an optimization framework for both narrowband and wideband systems based on a rigorous circuit-theoretic model. For narrowband systems, capacity maximization is formulated as a non-convex optimization problem, which is solved via a block coordinate ascent (BCA) framework. Because optimizing MA positions is challenging due to analytically intractable MC matrices, we develop a trust region method (TRM)-based algorithm that utilizes Sylvester equations to compute the derivatives of the inverse square roots of the MC matrices. We further consider wideband systems and formulate a sum-rate maximization problem. To find a unified set of MA positions that balances varying subcarrier conditions, the BCA framework and the TRM-based MA position optimization algorithm are extended to wideband systems. Simulation results demonstrate that exploiting MC effects in MA-MIMO systems yields significant performance gains in both narrowband and wideband systems under various channel conditions. These gains highlight the benefits of MC-induced superdirectivity and designable MC matrices.
\end{abstract} 

\begin{IEEEkeywords}
Movable antenna (MA), mutual coupling (MC), capacity, superdirectivity, OFDM, MIMO, optimization.
\end{IEEEkeywords}

\IEEEpeerreviewmaketitle
\section{Introduction}\label{sec:intro}
Multiple-input multiple-output (MIMO) is a key enabling technology of the sixth-generation (6G) wireless communications due to its ability to provide additional spatial degrees of freedom (DoFs), thereby improving both capacity and reliability~\cite{letaiefRoadmap6GAI2019,lin2024fundamentals}. A natural way to further exploit these DoFs is to scale up the antenna array, which has motivated the deployment of hundreds or even thousands of antennas at the base station~\cite{luTutorialNearFieldXLMIMO2024,yu2023adaptive}. However, simply increasing the number of antennas leads to substantial hardware complexity, power consumption, and signal processing costs. More importantly, conventional MIMO systems rely on fixed-position antennas and therefore lack the flexibility to adapt to changing propagation environments.

These limitations have motivated the development of reconfigurable antenna (RA) systems, such as movable antennas (MAs)~\cite{maMIMOCapacityCharacterization2024,zhu2025tutorial} and fluid antenna systems (FASs)~\cite{wongFluidAntennaSystems2021,new2025fluid,tang2025accurate}. MA systems can effectively mitigate deep fading caused by destructive multipath combinations. By moving antenna elements to positions where multipath components combine constructively, the channel quality can be significantly improved, thereby enhancing capacity~\cite{zhuModelingPerformanceAnalysis2024,zhu2025tutorial,liao2025joint}. Thus, MAs have been regarded as a disruptive enabling technology for 6G systems and have been applied to various scenarios, such as Internet of Things (IoT) networks~\cite{xiao2024throughput}, satellite communications~\cite{zhu2025dynamic}, and integrated sensing and communication (ISAC) systems~\cite{ma2025isac}. Antenna movement in MA systems gives rise to mutual coupling (MC), where the excitation of one antenna influences the signals radiated by the others. To mitigate this effect, existing studies on MA systems commonly impose a minimum antenna spacing of half a wavelength. However, this design suffers from several limitations:
\begin{enumerate*}
    \item MC effects still exist even under half-wavelength spacing constraints,
    \item the degrees of freedom (DoFs) for antenna movement are reduced, and
    \item the potential benefits of exploiting MC are forfeited.
\end{enumerate*}
Given these limitations, it is essential to explicitly account for MC effects in MA-MIMO systems.

\subsection{Related Works}
Extensive research has been devoted to exploiting the potential of MA-MIMO systems while neglecting MC effects. In~\cite{maMIMOCapacityCharacterization2024}, the authors first considered a point-to-point MA-MIMO system and formulated the capacity maximization problem as a non-concave optimization problem. A successive concave approximation (SCA) algorithm was proposed to solve this problem. Simulation results demonstrated that MA can achieve substantial capacity gains over conventional MIMO systems under various channel conditions. The benefits of MA systems in multi-user scenarios were further demonstrated in~\cite{zhu2024movable}. Specifically, a joint antenna position and receive combining optimization problem was formulated as a non-convex optimization problem, and both zero-forcing (ZF)-based and minimum mean square error (MMSE)-based multi-directional descent (MDD) algorithms were proposed to solve the problem. Simulation results showed that, owing to the interference mitigation capability of MA systems, the transmit power can be significantly reduced while maintaining the same rate performance. Recent studies have further investigated MA-MIMO systems in the near field~\cite{zhu2025nearfield}, under imperfect channel state information (CSI)~\cite{ma2025robust}, and in the presence of reconfiguration delays~\cite{wang2025throughput}. 

However, the above-mentioned studies typically assume flat fading channels, which are not valid for wideband systems. In such wideband scenarios, frequency-selective fading arises, which is commonly mitigated using orthogonal frequency division multiplexing (OFDM)~\cite{cimini1985analysis}. Since different subcarriers favor different antenna positions, optimization algorithms developed for narrowband systems cannot be applied to MA-enabled OFDM systems. To address this challenge, reference~\cite{zhu2024performance} proposed a channel modeling framework and a parallel greedy ascent (PGA) algorithm for joint subcarrier power allocation and MA position optimization in wideband MA-enabled single-input single-output (SISO) systems. For antenna position optimization in MA-enabled uplink multi-user single-input multiple-output (SIMO) systems, a particle swarm optimization (PSO) algorithm was developed in~\cite{irshadOptimizingMovableAntennas2025}. Simulation results in~\cite{zhu2024performance} and~\cite{irshadOptimizingMovableAntennas2025} demonstrated that MA can achieve superior wideband sum-rate performance over fixed-position antennas for various channel conditions and subcarrier numbers.

Despite the extensive studies on MA systems, MC effects have typically been neglected by imposing half-wavelength spacing constraints on antenna elements. However, rather than being a negligible artifact, MC is a fundamental physical phenomenon that has been investigated in the early stages of traditional MIMO research. In~\cite{wallaceMutualCouplingMIMO2004}, the authors modeled MIMO communication systems with MC using a circuit-theoretic framework and derived accurate expressions for MIMO channel capacity in the presence of MC effects. A more rigorous and comprehensive analytical framework was later developed in~\cite{ivrlacCircuitTheoryCommunication2010}, which explicitly accounted for both transmit and receive MC effects. As MIMO technology evolved, researchers gradually adopted half-wavelength antenna spacing as the standard configuration, as it reduces channel correlation and satisfies the Nyquist sampling theorem~\cite{tse2005fundamentals}. For isotropic antennas, such spacing effectively eliminates MC effects, whereas for non-isotropic antennas, MC effects become negligible. Consequently, MC is typically ignored in most existing MIMO studies.

Nevertheless, recent advancements in dense antenna deployments have necessitated the reincorporation of MC effects into MIMO systems~\cite{mezghaniReincorporatingCircuitTheory2024}. As researchers attempt to deploy increasingly large numbers of antennas within the same physical aperture, antennas are placed closer than half a wavelength apart and MC effects become more pronounced~\cite{gong2023holographic,yu2024bayes,qian2024including}. In conventional MIMO systems, MC is often regarded as detrimental because it degrades performance and complicates signal processing design. Accordingly, various decoupling techniques have been developed to mitigate MC~\cite{chenReviewMutualCoupling2018}. However, MC is not always harmful and can potentially be exploited to improve performance. In particular, MC-induced superdirectivity can achieve beam gains exceeding the number of antennas and thereby improve capacity in MIMO systems~\cite{morrisSuperdirectivityMIMOSystems2005}. Beyond superdirectivity, MC can also improve performance by designing MC matrices. The authors of~\cite{pizzo2025mutual} showed that properly designing the individual antenna patterns reshapes the MC matrices to better match the wireless channel statistics, thereby improving capacity. In MA-MIMO systems, antenna elements can move off the half-wavelength grid, giving rise to MC effects. MA-MIMO can leverage MC effects by enabling both superdirectivity effects and designable MC matrices. On the one hand, MC enables superdirectivity effects that enhance array directivity gains. On the other hand, antenna movement allows the MC matrices to be designed to better match the channel and improve capacity.

\subsection{Contributions}
This paper aims to investigate the gain achieved by leveraging MC effects in MA-MIMO systems. Leveraging MC effects in MA-MIMO systems poses an unprecedented challenge due to the analytically intractable MC matrices. To solve this challenge, we develop a novel optimization framework. The main contributions are summarized as follows:
\begin{itemize}
    \item We consider an MA-enabled point-to-point MIMO communication system that explicitly accounts for MC effects. By combining the field response channel model~\cite{maMIMOCapacityCharacterization2024}, which captures wireless propagation effects, with the circuit-based communication model~\cite{ivrlacCircuitTheoryCommunication2010}, which characterizes MC effects, we establish a unified framework for MA-MIMO systems with MC effects. To investigate the potential performance gains brought by MC, we formulate joint covariance matrix and antenna position optimization problems for both narrowband and wideband systems as non-concave optimization problems.
    \item To maximize the capacity in narrowband systems, we propose a block coordinate ascent (BCA)-based optimization framework. The subproblem of MA position optimization is particularly challenging due to the non-concavity and indefinite curvature of the objective function induced by the inverse square roots of the MC matrices. We therefore develop a trust region method (TRM)-based algorithm~\cite{yuan2015recent} for MA position optimization that requires only the first- and second-order derivatives of the objective function. Moreover, the first- and second-order derivatives of the inverse square roots of the MC matrices are derived in quasi-closed-form by solving Sylvester equations~\cite{higham2008functions}. Simulation results demonstrate that exploiting MC can achieve significant capacity gains in narrowband systems, even under strong line-of-sight (LoS) conditions and limited antenna movement ranges.
    \item We further investigate MA-MIMO systems in wideband scenarios with frequency-selective fading channels and formulate a sum-rate maximization problem that jointly optimizes covariance matrices and MA positions. In contrast to the narrowband case, the wideband MA position optimization is significantly more challenging. Since the optimal antenna positions vary across subcarriers, the optimization problem is highly coupled, as a unified set of MA positions must balance performance across all subcarriers. To tackle this challenge, we extend the BCA-based framework and develop a TRM-based algorithm that optimizes MA positions by jointly accounting for the varying channel conditions across all subcarriers. Simulation results show that the sum-rate gains achieved by exploiting MC become more pronounced as the number of subcarriers increases.
\end{itemize}
While concurrent studies~\cite{zhu2026mutual,xu2026directivity} have examined MC in MA-enabled systems, they primarily focus on its array and beam gain benefits. However, a rigorous understanding of MC in MA-MIMO systems and its impact on achievable capacity remains lacking. In this work, we bridge this gap by showing that capacity gains arise not only from superdirectivity but also from channel-aware design of the MC matrices.

\subsection{Organization and Notation}
The remainder of this paper is organized as follows. We present the system model and the circuit-based communication model in Section~\ref{sec:system_model}. In Section~\ref{sec:capacity_max_nb}, we formulate the narrowband capacity maximization problem and propose the BCA-based framework with the TRM-based algorithm to solve the problem. Section~\ref{sec:capacity_max_wb} formulates the wideband sum-rate maximization problem and extends the BCA-based framework and the TRM-based algorithm to wideband systems. Simulation results are provided in Section~\ref{sec:sim}. Finally, Section~\ref{sec:conclusion} concludes the paper.

\textit{Notation:} $a$, $\mathbf{a}$, and $\mathbf{A}$ denote a scalar, vector, and matrix, respectively. $[\mathbf{A}]_{m,n}$, $\mathbf{A}^\transpose$, $\mathbf{A}^\hermconj$, $\det(\mathbf{A})$, and $\trace(\mathbf{A})$ denote the $(m,n)$-th element, transpose, conjugate transpose, determinant, and trace of matrix $\mathbf{A}$, respectively. The imaginary unit is denoted by $j$. $\mathbf{Q} \succeq \mathbf{0}$ indicates that $\mathbf{Q}$ is positive semidefinite. Symbols $\partial(\cdot)$, $\Re\{\cdot\}$, $\lceil \cdot \rceil$, and $\mathsf{E}[\cdot]$ denote the partial derivative, real part, ceiling operator, and expectation operator, respectively. The sets of real numbers and complex numbers are denoted by $\mathbb{R}$ and $\mathbb{C}$, respectively. $\mathcal{CN}(\mathbf{0},\sigma^2\mathbf{I})$ denotes the circularly symmetric complex Gaussian (CSCG) distribution with zero mean and covariance matrix $\sigma^2\mathbf{I}$. $\mathcal{U}[a, b]$ represents the uniform distribution between $a$ and $b$.

\section{System Model}\label{sec:system_model}
\subsection{MA-MIMO System}
\begin{figure}[!t]
    \centering
    \includegraphics[width=0.45\textwidth]{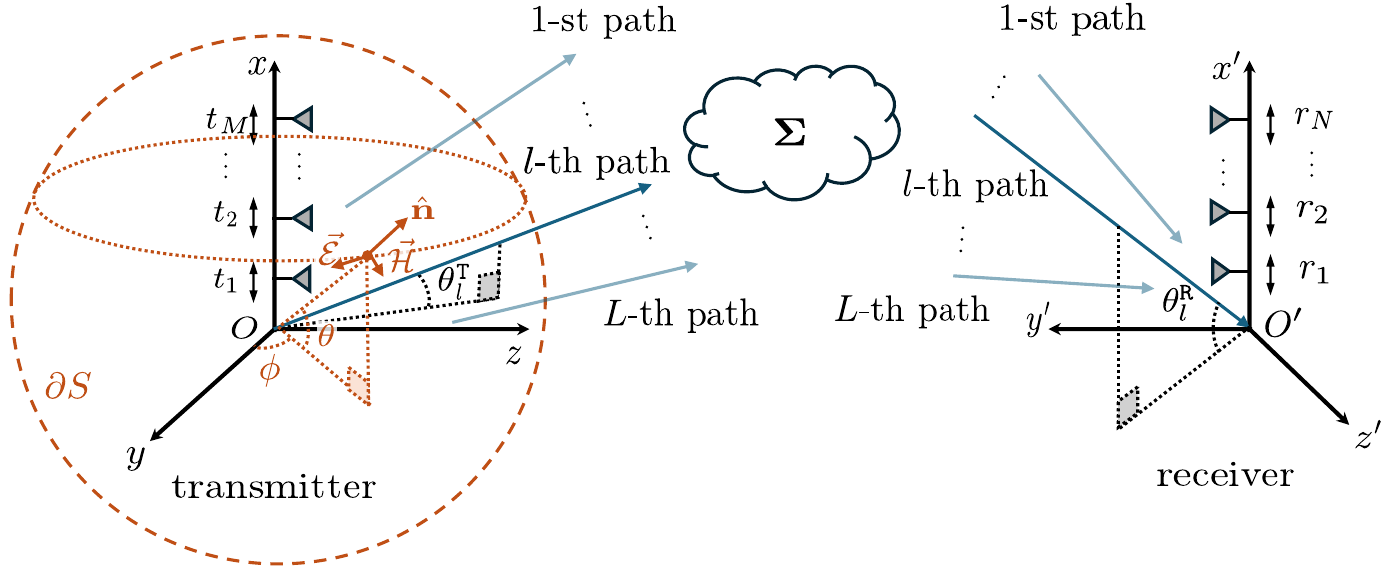}
    \caption{System model of a point-to-point MIMO communication system equipped with $M$ transmit MAs and $N$ receive MAs.}\label{fig:system_model}
\end{figure}
As illustrated in Fig.~\ref{fig:system_model}, we consider a point-to-point MIMO communication system equipped with $M$ transmit MAs and $N$ receive MAs. Local Cartesian coordinate systems, $(x,y,z)$ and $(x^\prime,y^\prime,z^\prime)$, are established at the transmitter and receiver, respectively. The transmit and receive MAs are constrained to move along the $x$-axis and the $x^\prime$-axis of their corresponding coordinate systems.\footnote{Although this work considers MC under one-dimensional MA movement for simplicity, the system model and proposed method can be extended to more general two- or three-dimensional MA-MIMO systems.} Therefore, the positions of the $m$-th ($m=1,2,\dots,M$) transmit MA and the $n$-th ($n=1,2,\dots,N$) receive MA are represented by the scalars $t_m$ and $r_n$, respectively. The position vectors of all transmit and receive MAs are denoted by $\mathbf{t} = [t_1, t_2, \dots, t_M]^\transpose \in \mathbb{R}^M$ and $\mathbf{r} = [r_1, r_2, \dots, r_N]^\transpose \in \mathbb{R}^N$, respectively.

We adopt the far-field field response model~\cite{maMIMOCapacityCharacterization2024} to characterize the wireless channel between the transmitter and receiver. The transmit and receive steering vectors are defined as

\begin{small}
    \vspace{-0.3cm}
    \begin{subequations}\label{eq:steering_vector_def}
        \begin{align}
            \mathbf{a}_{\mathtt{T}}(\mathbf{t},\theta) = [\mathrm{e}^{j k t_1 \sin\theta}, \mathrm{e}^{j k t_2 \sin\theta}, \dots, \mathrm{e}^{j k t_M \sin\theta}]^\transpose \in \mathbb{C}^M, \\
            \mathbf{a}_{\mathtt{R}}(\mathbf{r},\theta) = [\mathrm{e}^{j k r_1 \sin\theta}, \mathrm{e}^{j k r_2 \sin\theta}, \dots, \mathrm{e}^{j k r_N \sin\theta}]^\transpose \in \mathbb{C}^N,
        \end{align}
    \end{subequations}
\end{small}%
where $k = 2\pi/\lambda$, $\lambda$, and $\theta$ are the wavenumber, wavelength, and elevation angle, respectively. The number of multipath components is denoted by $L$.\footnote{We assume that the numbers of multipath components at the transmitter and receiver are identical. This assumption is consistent with the standard Saleh-Valenzuela (SV) channel model~\cite{salehStatisticalModelIndoor1987}.} We next define transmit and receive field response matrices (FRMs) by stacking the steering vectors:

\begin{small}
    \vspace{-0.3cm}
    \begin{subequations}\label{eq:frm_def}
        \begin{align}
            \mathbf{G}(\mathbf{t}) = [\mathbf{a}_{\mathtt{T}}(\mathbf{t},\theta_1^{\mathtt{T}}), \mathbf{a}_{\mathtt{T}}(\mathbf{t},\theta_2^{\mathtt{T}}), \dots, \mathbf{a}_{\mathtt{T}}(\mathbf{t},\theta_L^{\mathtt{T}})]^\transpose \in \mathbb{C}^{L \times M}, \\
            \mathbf{F}(\mathbf{r}) = [\mathbf{a}_{\mathtt{R}}(\mathbf{r},\theta_1^{\mathtt{R}}), \mathbf{a}_{\mathtt{R}}(\mathbf{r},\theta_2^{\mathtt{R}}), \dots, \mathbf{a}_{\mathtt{R}}(\mathbf{r},\theta_L^{\mathtt{R}})]^\transpose \in \mathbb{C}^{L \times N},
        \end{align}
    \end{subequations}
\end{small}%
respectively. Here, $\theta_l^{\mathtt{T}}$ and $\theta_l^{\mathtt{R}}$ $(l=1,2,\dots,L)$ represent the elevation angles of departure (AoDs) and angles of arrival (AoAs), respectively. The path response matrix (PRM) is further defined as $\mathbf{\Sigma}\triangleq \mathrm{diag}(b_1, b_2, \dots, b_L) \in\mathbb{C}^{L \times L}$, where $b_l$ denotes the channel response of the $l$-th multipath component. Therefore, the overall channel matrix between the transmitter and receiver is given by
\begin{equation}
    \tilde{\mathbf{H}}(\mathbf{t},\mathbf{r}) = \mathbf{F}^\hermconj(\mathbf{r}) \mathbf{\Sigma} \mathbf{G}(\mathbf{t}).
\end{equation}

In the existing MA literature, MC effects are generally ignored, and corresponding signal models are given by
\begin{equation}\label{eq:signal_model}
    \mathbf{y} = \tilde{\mathbf{H}}(\mathbf{t},\mathbf{r}) \mathbf{x} + \mathbf{n},
\end{equation}
where $\mathbf{y}\in\mathbb{C}^{N}$, $\mathbf{x}\in\mathbb{C}^{M}$, and $\mathbf{n}\sim\mathcal{CN}(\mathbf{0},\sigma^2\mathbf{I}_N)$ are the received signal, the transmitted signal, and additive white Gaussian noise (AWGN), respectively. The covariance matrix of the transmitted signal $\mathbf{x}$ is defined as $\mathbf{Q} \triangleq \E[\mathbf{x} \mathbf{x}^\hermconj] \in \mathbb{C}^{M\times M}$, with $\mathbf{Q} \succeq \mathbf{0}$. This model implicitly assumes that the transmitted power is defined as $P_{\mathtt{T}} = \E[\mathbf{x}^\hermconj \mathbf{x}]$. In practice, however, antenna movement alters the array's radiation pattern, causing the transmitted power to differ from the power radiated into free space. Consequently, the signal model~\eqref{eq:signal_model} may violate the principle of energy conservation.

From an electromagnetic perspective, the physically radiated power is given by the surface integral of the Poynting vector over the closed surface $\partial S$ enclosing the transmitter~\cite{morrisSuperdirectivityMIMOSystems2005}:
\begin{equation}\label{eq:radiated_power}
    P_{\mathtt{rad}} = \frac{1}{2}\oiint_{\partial S} \E[\Re\{\vec{\mathcal{E}} \times \vec{\mathcal{H}}^\star\}] \cdot \mathrm{d}\hat{\mathbf{n}},
\end{equation}
where $\vec{\mathcal{E}}$ and $\vec{\mathcal{H}}^\star$ denote the electric field and the conjugate of the magnetic field, respectively, and $\hat{\mathbf{n}}$ is the unit normal vector to the surface $\partial S$. If we omit scalar coefficients, assume isotropic antenna elements, and choose $\partial S$ as a sphere with radius $R_s$ centered at the origin of the transmit coordinate system, then~\eqref{eq:radiated_power} can be simplified to

\begin{small}
    \vspace{-0.3cm}
    \begin{align}\label{eq:radiated_power_simplified}
        P_{\mathtt{rad}} &= \E\Bigg[\mathbf{x}^\hermconj \underbrace{\left(\int_{0}^{\pi}\int_{0}^{2\pi}\mathbf{a}_{\mathtt{T}}(\mathbf{t},\theta)\mathbf{a}_{\mathtt{T}}^\hermconj(\mathbf{t},\theta) \sin\theta \mathrm{d}\phi \mathrm{d}\theta\right)}_{\mathbf{C}_{\mathtt{T}}(\mathbf{t})} \mathbf{x}\Bigg],
    \end{align}
\end{small}%
where $\phi$ represents the azimuth angle. The matrix $\mathbf{C}_{\mathtt{T}}(\mathbf{t})$ is referred to as the transmit \textbf{MC matrix} for isotropic antenna elements. Similarly, we define the receive MC matrix as $\mathbf{C}_{\mathtt{R}}(\mathbf{r})$. The elements of $\mathbf{C}_{\mathtt{T}}(\mathbf{t})$ and $\mathbf{C}_{\mathtt{R}}(\mathbf{r})$ are given by

\begin{small}
    \vspace{-0.3cm}
    \begin{subequations}\label{eq:mc_matrix_def}
        \begin{align}
            [\mathbf{C}_{\mathtt{T}}(\mathbf{t})]_{m,m^\prime} &= \mathrm{sinc}( k(t_m-t_{m^\prime})), \quad m,m^\prime = 1, 2, \dots, M, \\
            [\mathbf{C}_{\mathtt{R}}(\mathbf{r})]_{n,n^\prime} &= \mathrm{sinc}( k(r_n-r_{n^\prime})), \quad n,n^\prime = 1, 2, \dots, N,
        \end{align}
    \end{subequations}
\end{small}%
respectively, where $\mathrm{sinc}(x) = \sin(x)/x$. Obviously, the radiated power equals the transmitted power, $P_{\mathtt{rad}} = P_{\mathtt{T}}$, if and only if $\mathbf{C}_{\mathtt{T}}(\mathbf{t}) = \mathbf{I}_M$.
\begin{remark}
    In conventional MIMO systems, antenna elements are typically spaced with integer multiples of half a wavelength. With isotropic antennas, this configuration intentionally renders the MC matrices as identity matrices, i.e., $\mathbf{C}_{\mathtt{T}} = \mathbf{I}_M$ and $\mathbf{C}_{\mathtt{R}} = \mathbf{I}_N$, under which the MC effects are absent. In contrast, MA systems allow antenna elements to move off the half-wavelength grid, making MC inevitable. Existing studies on MA systems often ignore MC effects by enforcing a minimum antenna spacing of half a wavelength. Although the off-diagonal elements of $\mathbf{C}_{\mathtt{T}}(\mathbf{t})$ and $\mathbf{C}_{\mathtt{R}}(\mathbf{r})$ become relatively small under this constraint, MC still exists. Moreover, such simplifications overlook the influence of antenna movement on MC, leaving the impact of MC in MA systems unexplored.
\end{remark}

\subsection{Circuit-Based Communication Model}
\begin{figure}[!t]
    \centering
    \subfloat[Overall multiport circuit model.]{  
        \includegraphics[width=0.48\textwidth]{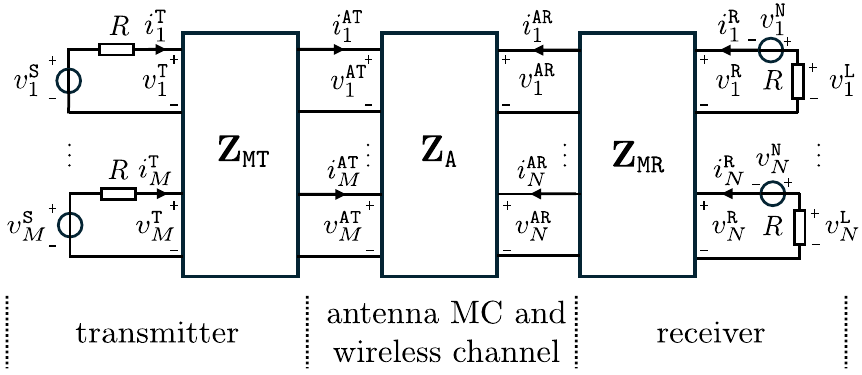}
        \label{fig:circuit_model}
    }

    \subfloat[Equivalent circuit observed at the input of $\mathbf{Z}_{\mathtt{MT}}$.]{  
        \includegraphics[width=0.212\textwidth]{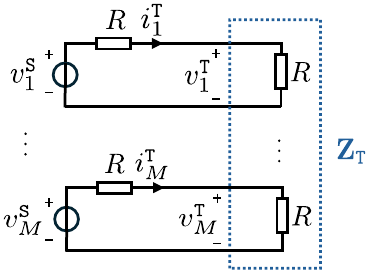}
        \label{fig:thevenin_tx}
    }\hspace{0.01\textwidth}
    \subfloat[Th\'evenin equivalent circuit at the output of $\mathbf{Z}_{\mathtt{MR}}$.]{  
        \includegraphics[width=0.228\textwidth]{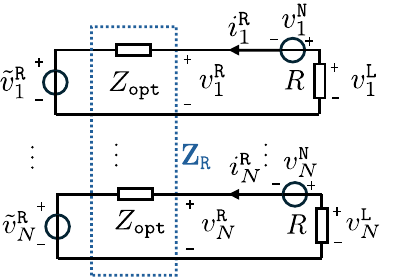}
        \label{fig:thevenin_rx}
    }
    \caption{Multiport circuit representation and equivalent models of the MA system with MC.}
\end{figure}

As discussed above, neglecting MC in MA systems leads to a fundamental mismatch between the radiated and transmitted powers. To resolve this physical inconsistency, a circuit-based communication framework is needed to ensure physical accuracy and electromagnetic compliance~\cite{ivrlacCircuitTheoryCommunication2010}. Specifically, under the assumption of perfect transmit power matching, receive noise matching, and isotropic background radiation, a simplified multiport circuit-based communication model is shown in Fig.~\subref*{fig:circuit_model}. The transmit and receive MC induced by antenna arrays are characterized by the impedance matrix $\mathbf{Z}_{\mathtt{A}}$:
\begin{equation}\label{eq:antenna_mc_model}
    \begin{bmatrix}
        \mathbf{v}_{\mathtt{AT}} \\
        \mathbf{v}_{\mathtt{AR}}
    \end{bmatrix}
    =
    \underbrace{\begin{bmatrix}
        \mathbf{Z}_{\mathtt{AT}} & \mathbf{0}_{M\times N} \\
        \mathbf{Z}_{\mathtt{ATR}} & \mathbf{Z}_{\mathtt{AR}}
    \end{bmatrix}}_{\mathbf{Z}_{\mathtt{A}}}
    \begin{bmatrix}
        \mathbf{i}_{\mathtt{AT}} \\
        \mathbf{i}_{\mathtt{AR}}
    \end{bmatrix},
\end{equation}
where $\mathbf{i}_{\mathtt{AT}} \triangleq [i_1^{\mathtt{AT}}, i_2^{\mathtt{AT}}, \dots, i_M^{\mathtt{AT}}]^\transpose$ and $\mathbf{v}_{\mathtt{AT}} \triangleq [v_1^{\mathtt{AT}}, v_2^{\mathtt{AT}}, \dots, v_M^{\mathtt{AT}}]^\transpose$ (and $\mathbf{i}_{\mathtt{AR}} \triangleq [i_1^{\mathtt{AR}}, i_2^{\mathtt{AR}}, \dots, i_N^{\mathtt{AR}}]^\transpose$ and $\mathbf{v}_{\mathtt{AR}} \triangleq [v_1^{\mathtt{AR}}, v_2^{\mathtt{AR}}, \dots, v_N^{\mathtt{AR}}]^\transpose$) are the currents and voltages at the input of the transmit MAs (and the output of the receive MAs), respectively. The matrices $\mathbf{Z}_{\mathtt{AT}}$ and $\mathbf{Z}_{\mathtt{AR}}$ denote the transmit and receive impedance matrices, respectively, where $\Re\{\mathbf{Z}_{\mathtt{AT}}\} = \mathbf{C}_{\mathtt{T}}(\mathbf{t})$ and $\Re\{\mathbf{Z}_{\mathtt{AR}}\} = \mathbf{C}_{\mathtt{R}}(\mathbf{r})$. The transimpedance matrix $\mathbf{Z}_{\mathtt{ATR}}$ characterizes the wireless channel response and is equivalent to the channel matrix $\tilde{\mathbf{H}}(\mathbf{t},\mathbf{r})$, up to constant scaling factors. Under the unilateral approximation~\cite{ivrlacCircuitTheoryCommunication2010}, the contribution of the receive currents $\mathbf{i}_{\mathtt{AR}}$ to the transmit voltages $\mathbf{v}_{\mathtt{AT}}$ is neglected, since the round-trip signal is sufficiently weak.

To ensure efficient power transfer from the signal generator $\mathbf{v}_{\mathtt{S}} \triangleq [v^{\mathtt{S}}_1, v^{\mathtt{S}}_2, \dots, v^{\mathtt{S}}_M]^\transpose$ to the MAs, a lossless impedance matching network $\mathbf{Z}_{\mathtt{MT}} \in \mathbb{C}^{2M\times 2M}$ is employed at the transmitter, as given by~\cite[Eq.~(103)]{ivrlacCircuitTheoryCommunication2010}. Under the unilateral approximation, the circuit seen at the input of $\mathbf{Z}_{\mathtt{MT}}$ reduces to a passive network with impedance $\mathbf{Z}_{\mathtt{T}} = R \mathbf{I}_M$. Here, $R$ denotes the transmit internal resistance. The equivalent circuit is illustrated in Fig.~\subref*{fig:thevenin_tx}. Consequently, the transmit voltages at the input of $\mathbf{Z}_{\mathtt{MT}}$ satisfy
\begin{equation}\label{eq:thevenin_tx_eq}
    \mathbf{v}_{\mathtt{T}} = \tfrac{1}{2} \mathbf{v}_{\mathtt{S}},
\end{equation}
where $\mathbf{v}_{\mathtt{T}} \triangleq [v^{\mathtt{T}}_1, v^{\mathtt{T}}_2, \dots, v^{\mathtt{T}}_M]^\transpose$. To maximize the received signal-to-noise ratio (SNR), a lossless impedance matching network $\mathbf{Z}_{\mathtt{MR}} \in \mathbb{C}^{2N\times 2N}$ is employed at the receiver, whose expression is provided in~\cite[Eq.~(84)]{ivrlacCircuitTheoryCommunication2010}. Since the circuit observed at the output of $\mathbf{Z}_{\mathtt{MR}}$ is active due to the source $\mathbf{v}_{\mathtt{S}}$, it can be transformed into an equivalent circuit via Th\'evenin's theorem~\cite{ivrlacCircuitTheoryCommunication2010}, as shown in Fig.~\subref*{fig:thevenin_rx}. The equivalent impedance is $\mathbf{Z}_{\mathtt{R}} = Z_{\mathtt{opt}}\mathbf{I}_N$, where $Z_{\mathtt{opt}}$ denotes the optimal output impedance under perfect noise matching. Based on the model in Fig.~\subref*{fig:thevenin_rx}, we can easily obtain
\begin{equation}\label{eq:thevenin_rx_eq}
    \mathbf{v}_{\mathtt{L}} = \tfrac{R}{Z_{\mathtt{opt}} + R} \left( \tilde{\mathbf{v}}_{\mathtt{R}} + \mathbf{v}_{\mathtt{N}} \right),
\end{equation}
where $\tilde{\mathbf{v}}_{\mathtt{R}} \triangleq [\tilde{v}^{\mathtt{R}}_1, \dots, \tilde{v}^{\mathtt{R}}_N]^\transpose$ and $\mathbf{v}_{\mathtt{N}} \triangleq [v^{\mathtt{N}}_1, \dots, v^{\mathtt{N}}_N]^\transpose$ denote the Th\'evenin equivalent voltage and the noise voltage, respectively. The equivalent voltage $\tilde{\mathbf{v}}_{\mathtt{R}}$ is obtained by setting $\mathbf{i}_{\mathtt{R}} = \mathbf{0}$ in the original circuit. Combining~\eqref{eq:antenna_mc_model}, \eqref{eq:thevenin_tx_eq}, and~\cite[Eqs.~(84) and (103)]{ivrlacCircuitTheoryCommunication2010}, we have
\begin{equation}\label{eq:thevenin_rx_eq_final}
    \tilde{\mathbf{v}}_{\mathtt{R}} = -\sqrt{\tfrac{\Re\{Z_{\mathtt{opt}}\}}{4R}}
    \mathbf{C}_{\mathtt{R}}^{-1/2}(\mathbf{r})
    \tilde{\mathbf{H}}(\mathbf{t},\mathbf{r})
    \mathbf{C}_{\mathtt{T}}^{-1/2}(\mathbf{t})
    \mathbf{v}_{\mathtt{S}}.
\end{equation}

Finally, substituting~\eqref{eq:thevenin_rx_eq_final} into~\eqref{eq:thevenin_rx_eq} yields the input-output relation of the system:
\begin{equation}\label{eq:system_model_mc}
    \underbrace{\tfrac{1}{2\sqrt{R}}\mathbf{v}_{\mathtt{L}}}_{\mathbf{y}}
    =
    \underbrace{\mathbf{C}_{\mathtt{R}}^{-1/2}(\mathbf{r})
    \tilde{\mathbf{H}}(\mathbf{t},\mathbf{r})
    \mathbf{C}_{\mathtt{T}}^{-1/2}(\mathbf{t})}_{\mathbf{H}(\mathbf{t},\mathbf{r})}
    \underbrace{\tfrac{1}{2\sqrt{R}}\mathbf{v}_{\mathtt{S}}}_{\mathbf{x}}
    +
    \underbrace{\tfrac{-1}{\sqrt{\Re\{Z_{\mathtt{opt}}\}}}\mathbf{v}_{\mathtt{N}}}_{\mathbf{n}},
\end{equation}
where $\mathbf{v}_{\mathtt{L}} \triangleq [v^{\mathtt{L}}_1, v^{\mathtt{L}}_2, \dots, v^{\mathtt{L}}_N]^\transpose$, $\mathbf{v}_{\mathtt{N}} \triangleq [v^{\mathtt{N}}_1, v^{\mathtt{N}}_2, \dots, v^{\mathtt{N}}_N]^\transpose$, and $\mathbf{H}(\mathbf{t},\mathbf{r})$ denote the load voltage vector, noise voltage vector, and equivalent channel matrix, respectively. $\mathbf{x}$, $\mathbf{y}$, and $\mathbf{n}$ represent the transmitted signal, received signal, and noise, respectively, consistent with~\eqref{eq:signal_model}. The radiated power $P_{\mathtt{rad}}$ now equals the transmitted power $P_{\mathtt{T}}$ with MC taken into account:
\begin{equation}
    \setlength{\abovedisplayskip}{8pt}
    \setlength{\belowdisplayskip}{8pt}
    P_{\mathtt{rad}}
    = \E[\mathbf{i}_{\mathtt{AT}}^\hermconj \mathbf{C}_{\mathtt{T}}(\mathbf{t}) \mathbf{i}_{\mathtt{AT}}]
    = \E\!\left[\tfrac{1}{4R}\mathbf{v}_{\mathtt{S}}^\hermconj \mathbf{v}_{\mathtt{S}}\right]
    = \E[\mathbf{x}^\hermconj \mathbf{x}]
    = P_{\mathtt{T}},
\end{equation}
where the second equality follows from~\cite[Eq.~(103)]{ivrlacCircuitTheoryCommunication2010}. With the input-output relation specified in~\eqref{eq:system_model_mc}, we can now investigate the impact of MC effects on the capacity of the MA system in both narrowband and wideband regimes.

\section{MC-Aware Capacity Maximization for Narrowband Systems}\label{sec:capacity_max_nb}
In this section, we focus on the narrowband communication system and assume that the channel experiences flat fading. Under this assumption, the channel bandwidth is much smaller than the coherence bandwidth and the channel can therefore be regarded as a flat channel. The system thus operates over a single carrier, and its performance is evaluated by the capacity normalized by the carrier bandwidth. Based on this assumption, we first formulate a capacity maximization problem that incorporates MC effects, and then develop a BCA-based optimization algorithm to solve the problem.

\subsection{Problem Formulation}
We aim to investigate the impact of MC on the theoretical capacity limit of MA-enabled narrowband MIMO systems based on the circuit-based signal model~\eqref{eq:system_model_mc}. In narrowband systems, the channel response can be regarded as flat within the communication bandwidth, and the capacity is therefore independent of the operating frequency:
\begin{equation}\label{eq:capacity}
    C(\mathbf{t},\mathbf{r}) = \max_{\substack{\mathbf{Q}: \mathbf{Q} \succeq \mathbf{0} \\ \trace(\mathbf{Q})\leq P_{\max}}} \log \det\left(\mathbf{I}_N + \tfrac{1}{\sigma^2}
        \mathbf{H}(\mathbf{t},\mathbf{r}) \mathbf{Q} \mathbf{H}^\hermconj(\mathbf{t},\mathbf{r})\right).
\end{equation}
Unlike conventional MIMO systems, the capacity $C(\mathbf{t},\mathbf{r})$ is a function of transmit and receive MA positions $\mathbf{t}$ and $\mathbf{r}$. Thus, we formulate the capacity maximization problem as
\begin{subequations}\label{eq:se_max_problem}
    \begin{align}
        \max_{\mathbf{Q},\mathbf{t},\mathbf{r}} \quad &
        \log \det\!\left(\mathbf{I}_N + \tfrac{1}{\sigma^2}
        \mathbf{H}(\mathbf{t},\mathbf{r}) \mathbf{Q} \mathbf{H}^\hermconj(\mathbf{t},\mathbf{r})\right), \label{eq:se_max_obj}\\
        \text{s.t.} \quad &
        \mathbf{Q} \succeq \mathbf{0}, \label{eq:se_max_q_psd}\\
        &
        \trace(\mathbf{Q}) \leq P_{\max}, \label{eq:se_max_power}\\
        &
        0 \leq t_m \leq D_{\mathtt{T}}, \quad m = 1,2,\dots,M, \label{eq:se_max_t_range}\\
        &
        0 \leq r_n \leq D_{\mathtt{R}}, \quad n = 1,2,\dots,N, \label{eq:se_max_r_range}\\
        &
        t_m - t_{m-1} \geq d_{\min}, \quad m = 2,3,\dots,M, \label{eq:se_max_t_spacing}\\
        &
        r_n - r_{n-1} \geq d_{\min}, \quad n = 2,3,\dots,N. \label{eq:se_max_r_spacing}
    \end{align}
\end{subequations}
Here, constraint~\eqref{eq:se_max_q_psd} enforces the positive semidefiniteness of $\mathbf{Q}$, and~\eqref{eq:se_max_power} limits the total transmit power to $P_{\max}$. Constraints~\eqref{eq:se_max_t_range} and~\eqref{eq:se_max_r_range} restrict the transmit and receive MAs within their respective movement ranges. Constraints~\eqref{eq:se_max_t_spacing} and~\eqref{eq:se_max_r_spacing} ensure a minimum spacing between adjacent MAs.\footnote{In MA systems without MC~\cite{maMIMOCapacityCharacterization2024,zhuModelingPerformanceAnalysis2024}, the spacing constraints~\eqref{eq:se_max_t_spacing} and~\eqref{eq:se_max_r_spacing} are typically imposed to mitigate MC among antenna elements, where the minimum spacing $d_{\min}$ is usually set exactly as $\lambda/2$. Although MC effects are explicitly considered in this paper, such constraints remain necessary to prevent physical collisions between MAs with much smaller $d_{\min}$ than $\lambda/2$. Moreover, it suffices to assume that the ordering of the MAs is preserved, i.e., $t_1<t_2<\dots<t_M$ and $r_1<r_2<\dots<r_N$, since linear MA arrays are considered in this paper.}

Solving the optimization problem~\eqref{eq:se_max_problem} is the key to fully exploiting MC in MA systems. However, the problem is challenging to solve due to the non-concavity of its objective function. Compared with the capacity maximization problem that ignores MC~\cite{maMIMOCapacityCharacterization2024}, the presence of the additional MC matrices $\mathbf{C}_{\mathtt{T}}(\mathbf{t})$ and $\mathbf{C}_{\mathtt{R}}(\mathbf{r})$ makes the problem more challenging. In particular, the inverse square roots $\mathbf{C}^{-1/2}_{\mathtt{T}}(\mathbf{t})$ and $\mathbf{C}^{-1/2}_{\mathtt{R}}(\mathbf{r})$ lack closed-form expressions, making gradient derivations analytically intractable. In the remainder of this section, we propose a BCA-based algorithm to iteratively solve problem~\eqref{eq:se_max_problem}. In each iteration, each set of parameters is optimized in an alternating manner with the others fixed. Specifically, the transmit covariance matrix $\mathbf{Q}$ is optimized using the water-filling algorithm~\cite{tse2005fundamentals}, and the MA positions $\mathbf{t}$ and $\mathbf{r}$ are optimized using the TRM~\cite{yuan2015recent}.

\subsection{Update of $\mathbf{Q}$}\label{subsec:q}
In this step, we optimize the transmit covariance matrix $\mathbf{Q}$ while keeping the MA positions $\mathbf{t}$ and $\mathbf{r}$ fixed. The optimization problem~\eqref{eq:se_max_problem} can be reformulated as
\begin{subequations}\label{eq:q_problem}
    \begin{align}
        \max_{\mathbf{Q}} \quad &\log \det\left(\mathbf{I}_N + \tfrac{1}{\sigma^2}
        \mathbf{H}(\overline{\mathbf{t}},\overline{\mathbf{r}}) \mathbf{Q} \mathbf{H}^\hermconj(\overline{\mathbf{t}},\overline{\mathbf{r}})\right),\\
        \text{s.t.}\quad &\mathbf{Q} \succeq \mathbf{0}, \\
        &\trace(\mathbf{Q}) \leq P_{\max}.
    \end{align}
\end{subequations}
This subproblem is concave and can be efficiently solved using the water-filling algorithm~\cite{tse2005fundamentals}. Let the singular value decomposition (SVD) of $\mathbf{H}(\overline{\mathbf{t}},\overline{\mathbf{r}})$ be expressed as $\mathbf{H}(\overline{\mathbf{t}},\overline{\mathbf{r}}) = \overline{\mathbf{U}} \overline{\mathbf{\Lambda}} \overline{\mathbf{V}}^\hermconj$, where $\overline{\mathbf{\Lambda}} = \mathrm{diag}(\lambda_1, \lambda_2, \dots, \lambda_\Gamma) \in \mathbb{R}^{\Gamma\times \Gamma}$, $\overline{\mathbf{U}} \in \mathbb{C}^{N\times \Gamma}$, and $\overline{\mathbf{V}} \in \mathbb{C}^{M\times \Gamma}$ denote the singular values, left singular vectors, and right singular vectors of $\mathbf{H}(\overline{\mathbf{t}},\overline{\mathbf{r}})$, respectively. Here, $\overline{\mathbf{t}}$ and $\overline{\mathbf{r}}$ represent the transmit and receive MA positions obtained in the previous iteration, and $\Gamma$ denotes the rank of $\mathbf{H}(\overline{\mathbf{t}},\overline{\mathbf{r}})$.

According to the water-filling principle, the transmit power allocated to the $\gamma$-th eigenchannel is
\begin{equation*}
    P_\gamma = \max\left(0,\, \mu - \tfrac{\sigma^2}{\lambda_\gamma^2}\right), \qquad \gamma = 1,2,\dots,\Gamma,
\end{equation*}
where $\mu$ is the water level chosen via bisection search to satisfy $\sum_{\gamma=1}^{\Gamma} P_\gamma = P_{\max}$. Thus, the optimal transmit covariance matrix is given by
\begin{equation}\label{eq:q_opt}
    \mathbf{Q} = \overline{\mathbf{V}}\, \mathrm{diag}(P_1, P_2, \dots, P_\Gamma)\, \overline{\mathbf{V}}^\hermconj.
\end{equation}

\subsection{Update of $t_m$}\label{subsec:t_m}
In this step, we optimize the $m$-th transmit MA position $t_m$ while keeping the transmit covariance matrix $\mathbf{Q}$, the receive MA positions $\mathbf{r}$, and the other transmit MA positions $t_{m^\prime}\ (m^\prime \neq m)$ fixed. The optimization problem~\eqref{eq:se_max_problem} can be reformulated as
\begin{equation}\label{eq:t_m_problem}
    \setlength{\abovedisplayskip}{8pt}
    \setlength{\belowdisplayskip}{8pt}
    \max_{t_m}\ h_\mathtt{T}(t_m), \quad \text{s.t.}\ \overline{t}_{m-1}+d_{\min}\leq t_m \leq \overline{t}_{m+1}-d_{\min},
\end{equation}
where we define $\overline{t}_0 = -d_{\min}$ and $\overline{t}_{M+1} = D_{\mathtt{T}}+d_{\min}$ for convenience. The objective function $h_\mathtt{T}(t_m)$ equals $C(\mathbf{t},\mathbf{r})$ in~\eqref{eq:capacity} with only $t_m$ being the variable, and it remains non-concave. Determining update step sizes that ensure monotonic increases in $h_\mathtt{T}(t_m)$ is challenging, as its curvature upper bound cannot be explicitly derived. Therefore, we adopt the TRM~\cite{yuan2015recent} to solve problem~\eqref{eq:t_m_problem}, as it guarantees convergence using only the first- and second-order partial derivatives of $h_\mathtt{T}(t_m)$ and remains effective even with indefinite curvature.

\begin{figure}[tb]
    \centering
    \includegraphics[width=0.4\textwidth]{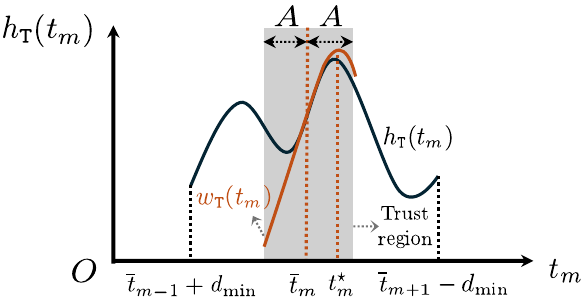}
    \caption{TRM for updating $t_m$.}
    \label{fig:trust_region}
\end{figure}

Fig.~\ref{fig:trust_region} illustrates the TRM for updating $t_m$. The trust region (TR) is defined as the set of candidate values centered at $\overline{t}_m$ with a radius of $A$. Within the intersection of the TR and the feasible set $\mathcal{T}_m$ defined by the constraint of~\eqref{eq:t_m_problem}, the objective function $h_\mathtt{T}(t_m)$ is locally approximated by a quadratic function $w_{\mathtt{T}}(t_m)$:

\begin{small}
    \vspace{-0.3cm}
    \begin{equation}\label{eq:w_t_m}
        w_{\mathtt{T}}(t_m)
        = \tfrac{1}{2} h^{\prime\prime}_\mathtt{T}(\overline{t}_m)(t_m - \overline{t}_m)^2
        + h^{\prime}_\mathtt{T}(\overline{t}_m)(t_m - \overline{t}_m)
        + h_\mathtt{T}(\overline{t}_m),
    \end{equation}
\end{small}%
where the functions $h_\mathtt{T}^{\prime}$ and $h_\mathtt{T}^{\prime\prime}$ denote the first- and second-order derivatives of $h_\mathtt{T}(t_m)$, respectively. The computation of these derivatives will be detailed later. The feasible set $\mathcal{T}_m$ is defined as
\begin{equation}\label{eq:t_m_feasible_set}
    \mathcal{T}_m \triangleq 
    [\max(\overline{t}_m - A,\overline{t}_{m-1} + d_{\min}),
    \min(\overline{t}_m + A,\overline{t}_{m+1} - d_{\min})].
\end{equation}
Since $w_{\mathtt{T}}(t_m)$ is quadratic, its optimal solution is given by
\begin{equation}\label{eq:t_m_star}
    t_m^\star = \mathrm{Proj}_{\mathcal{T}_m}\!\left(
    \overline{t}_m - 
    h^{\prime}_\mathtt{T}(\overline{t}_m)/h^{\prime\prime}_\mathtt{T}(\overline{t}_m)
    \right),
\end{equation}
where $\mathrm{Proj}_{\mathcal{T}_m}(\cdot)$ denotes the projection operator onto the set $\mathcal{T}_m$. If $w_{\mathtt{T}}(t_m)$ provides a good local approximation of $h_\mathtt{T}(t_m)$, the optimum $t_m^\star$ of $w_{\mathtt{T}}(t_m)$ leads to an improvement in $h_\mathtt{T}(t_m)$. The approximation quality is evaluated by the ratio
\begin{equation}\label{eq:rho}
    \varrho = 
    \frac{h_\mathtt{T}(\overline{t}_m) - h_\mathtt{T}(t_m^\star)}
         {w_\mathtt{T}(\overline{t}_m) - w_{\mathtt{T}}(t_m^\star)}.
\end{equation}
We define two constant thresholds, $\varrho_1$ and $\varrho_2$, satisfying $0 < \varrho_1 < \varrho_2 < 1$, to evaluate the approximation quality. If $\varrho > \varrho_2$, $w_{\mathtt{T}}(t_m)$ is considered a good local approximation of $h_\mathtt{T}(t_m)$. In this case, the update is accepted by setting $t_m = t_m^\star$. Moreover, if $t_m^\star$ lies on the boundary of the TR, the radius $A$ increases to $\nu_1 A$ for the next iteration. If $\varrho_1 < \varrho \leq \varrho_2$, the update is accepted while keeping $A$ unchanged. If $\varrho \leq \varrho_1$, the update is rejected, and the radius is reduced to $A/\nu_2$.

The key to applying the TRM lies in computing the first- and second-order derivatives of $h_\mathtt{T}(t_m)$. Using the matrix chain rule~\cite{petersen2008matrix}, we obtain the first- and second-order derivatives of $h_\mathtt{T}(t_m)$ as
\begin{align}
    h^{\prime}_\mathtt{T}(t_m) &= \tfrac2{\sigma^2}\Re\left\{\trace\left[\mathbf{\Phi} \tfrac{\partial \mathbf{H}}{\partial t_m} \overline{\mathbf{Q}}\mathbf{H}^\hermconj\right] \right\},\label{eq:h_prime_t_m}\\
    h^{\prime\prime}_\mathtt{T}(t_m) &= \tfrac2{\sigma^2}\Re\left\{\trace\left[\mathbf{\Phi}\left(\tfrac{\partial^2 \mathbf{H}}{\partial t_m^2} \overline{\mathbf{Q}}\mathbf{H}^\hermconj + \tfrac{\partial \mathbf{H}}{\partial t_m} \overline{\mathbf{Q}} \tfrac{\partial \mathbf{H}^\hermconj}{\partial t_m}\right)\right. \right.\nonumber \\
    &\hspace{3em}-\sigma^{-2}\mathbf{\Phi}\tfrac{\partial \mathbf{H}}{\partial t_m} \overline{\mathbf{Q}}\mathbf{H}^\hermconj \mathbf{\Phi}\tfrac{\partial \mathbf{H}}{\partial t_m}\overline{\mathbf{Q}}\mathbf{H}^\hermconj \nonumber \\
    &\hspace{3em}-\left.\left.\sigma^{-2} \mathbf{\Phi}\tfrac{\partial \mathbf{H}}{\partial t_m} \overline{\mathbf{Q}}\mathbf{H}^\hermconj \mathbf{\Phi}\mathbf{H}\overline{\mathbf{Q}}\tfrac{\partial \mathbf{H}^\hermconj}{\partial t_m}\right]\right\},\label{eq:h_prime2_t_m}
\end{align}
where $\mathbf{\Phi} \triangleq \left(\mathbf{I}_N + \sigma^{-2}\mathbf{H}\overline{\mathbf{Q}}\mathbf{H}^\hermconj\right)^{-1}$ and $\overline{\mathbf{Q}}$ denotes the temporarily optimized $\mathbf{Q}$. The first- and second-order derivatives of $\mathbf{H}$ with respect to (w.r.t.) $t_m$ are given by

\begin{small}
\vspace{-0.3cm}
\begin{align}
    \frac{\partial \mathbf{H}}{\partial t_m} &= \overline{\mathbf{C}}_{\mathtt{R}}^{-\frac12} \overline{\mathbf{F}}^\hermconj \mathbf{\Sigma} \Big[\frac{\partial \mathbf{G}}{\partial t_m} \mathbf{C}_{\mathtt{T}}^{-\frac12} + \mathbf{G} \frac{\partial \mathbf{C}_{\mathtt{T}}^{-1/2}}{\partial t_m}\Big], \label{eq:partial_h_t_m}\\
    \frac{\partial^2 \mathbf{H}}{\partial t_m^2} &= \overline{\mathbf{C}}_{\mathtt{R}}^{-\frac12} \overline{\mathbf{F}}^\hermconj \mathbf{\Sigma}\Big[\frac{\partial^2 \mathbf{G}}{\partial t_m^2} \mathbf{C}_{\mathtt{T}}^{-\frac12} + 2\frac{\partial \mathbf{G}}{\partial t_m} \frac{\partial \mathbf{C}_{\mathtt{T}}^{-1/2}}{\partial t_m} + \mathbf{G} \frac{\partial^2 \mathbf{C}_{\mathtt{T}}^{-1/2}}{\partial t_m^2}\Big], \label{eq:partial2_h_t_m}
\end{align}
\end{small}%
respectively. Here $\overline{\mathbf{C}}_{\mathtt{R}}$ and $\overline{\mathbf{F}}$ are taken from the previous iteration. In~\eqref{eq:partial_h_t_m} and~\eqref{eq:partial2_h_t_m}, the elements of the first- and second-order derivatives of $\mathbf{G}$ w.r.t. $t_m$ are
\begin{align}
    \Big[\frac{\partial \mathbf{G}}{\partial t_m}\Big]_{l, m^{\prime}} &= \left\{
        \begin{aligned}
            &j k\sin\theta_l \mathrm{e}^{j k t_{m^\prime}\sin\theta_l}, &  m^\prime = m,\\
            &0, & m^\prime \neq m,
        \end{aligned}\right.\label{eq:partial_g_t_m}\\
    \Big[\frac{\partial^2 \mathbf{G}}{\partial t_m^2}\Big]_{l, m^{\prime}} &= \left\{
        \begin{aligned}
            &- k^2\sin^2\theta_l \mathrm{e}^{j k t_{m^\prime}\sin\theta_l}, &  m^\prime = m,\\
            &0, & m^\prime \neq m,
        \end{aligned}\right.\label{eq:partial2_g_t_m}
\end{align}
respectively. Note that in~\eqref{eq:partial_h_t_m} and~\eqref{eq:partial2_h_t_m}, the first- and second-order partial derivatives of $\mathbf{C}_{\mathtt{T}}^{-1/2}$ w.r.t. $t_m$ are analytically intractable. Instead, we first construct equations characterizing the partial derivatives of $\mathbf{C}_{\mathtt{T}}^{-1/2}$ and then obtain their exact values by solving these equations. We begin by deriving the equation characterizing $\frac{\partial \mathbf{C}_{\mathtt{T}}^{-1/2}}{\partial t_m}$. By taking the partial derivative on both sides of the identity $\mathbf{C}_{\mathtt{T}}^{-1/2} \mathbf{C}_{\mathtt{T}} \mathbf{C}_{\mathtt{T}}^{-1/2} = \mathbf{I}_M$ w.r.t. $t_m$, we have
\begin{equation}\label{eq:sylvester_equation_1}
    \frac{\partial \mathbf{C}_{\mathtt{T}}^{-1/2}}{\partial t_m} \mathbf{C}_{\mathtt{T}}^{\frac12}
    + \mathbf{C}_{\mathtt{T}}^{\frac12} \frac{\partial \mathbf{C}_{\mathtt{T}}^{-1/2}}{\partial t_m}
    = -\mathbf{C}_{\mathtt{T}}^{-\frac12} \frac{\partial \mathbf{C}_{\mathtt{T}}}{\partial t_m} \mathbf{C}_{\mathtt{T}}^{-\frac12}.
\end{equation}
The elements of $\frac{\partial \mathbf{C}_{\mathtt{T}}}{\partial t_m}$ are given by
\begin{small}
\begin{align}\label{eq:partial_c_t}
    &\Big[\frac{\partial \mathbf{C}_{\mathtt{T}}}{\partial t_m}\Big]_{m_1, m_2} \nonumber \\
    =&
    \begin{cases}
        \dfrac{\cos k(t_{m_1}-t_{m_2})}{t_{m_1}-t_{m_2}}
        - \dfrac{\sin k(t_{m_1}-t_{m_2})}{ k(t_{m_1}-t_{m_2})^2}, & m_1 \neq m_2 = m,\\
        \Big[\dfrac{\partial \mathbf{C}_{\mathtt{T}}}{\partial t_m}\Big]_{m_2, m_1}, & m_2 \neq m_1 = m,\\
        0, & \text{otherwise}.
    \end{cases}
\end{align}
\end{small}%
Similarly, the equation characterizing $\frac{\partial^2 \mathbf{C}_{\mathtt{T}}^{-1/2}}{\partial t_m^2}$ can be obtained by taking the partial derivative on both sides of~\eqref{eq:sylvester_equation_1} w.r.t. $t_m$:
\begin{align}\label{eq:sylvester_equation_2}
    \frac{\partial^2 \mathbf{C}_{\mathtt{T}}^{-1/2}}{\partial t_m^2} \mathbf{C}_{\mathtt{T}}^{\frac12}
    + \mathbf{C}_{\mathtt{T}}^{\frac12} \frac{\partial^2 \mathbf{C}_{\mathtt{T}}^{-1/2}}{\partial t_m^2}
    = &-\mathbf{C}_{\mathtt{T}}^{-\frac12} \frac{\partial^2 \mathbf{C}_{\mathtt{T}}}{\partial t_m^2} \mathbf{C}_{\mathtt{T}}^{-\frac12} \\ \nonumber
    - \mathbf{C}_{\mathtt{T}}^{-\frac12} \frac{\partial \mathbf{C}_{\mathtt{T}}}{\partial t_m} \frac{\partial \mathbf{C}_{\mathtt{T}}^{-1/2}}{\partial t_m}
    &- \frac{\partial \mathbf{C}_{\mathtt{T}}^{-1/2}}{\partial t_m} \frac{\partial \mathbf{C}_{\mathtt{T}}}{\partial t_m} \mathbf{C}_{\mathtt{T}}^{-\frac12} \\ \nonumber
    - \frac{\partial \mathbf{C}_{\mathtt{T}}^{1/2}}{\partial t_m} \frac{\partial \mathbf{C}_{\mathtt{T}}^{-1/2}}{\partial t_m}
    &- \frac{\partial \mathbf{C}_{\mathtt{T}}^{-1/2}}{\partial t_m} \frac{\partial \mathbf{C}_{\mathtt{T}}^{1/2}}{\partial t_m},
\end{align}
where the elements of $\frac{\partial^2 \mathbf{C}_{\mathtt{T}}}{\partial t_m^2}$ are given by
\begin{small} 
\begin{align} \label{eq:partial2_c_t}
    &\Big[\frac{\partial^2 \mathbf{C}_{\mathtt{T}}}{\partial t_m^2}\Big]_{m_1, m_2} \nonumber \\ 
    =& \begin{cases} 
        - k\frac{\sin k(t_{m_1}-t_{m_2})}{t_{m_1}-t_{m_2}} - 2\frac{\cos k(t_{m_1}-t_{m_2})}{ k(t_{m_1}-t_{m_2})^2} 
        &+ 2\frac{\sin k(t_{m_1}-t_{m_2})}{ k(t_{m_1}-t_{m_2})^3}, \\ 
        & m_1 \neq m_2 = m,\\ \Big[\dfrac{\partial^2 \mathbf{C}_{\mathtt{T}}}{\partial t_m^2}\Big]_{m_2, m_1}, & m_2 \neq m_1 = m,\\ 
        0, & \text{otherwise}.
    \end{cases} 
\end{align} 
\end{small}%
The term $\frac{\partial \mathbf{C}_{\mathtt{T}}^{1/2}}{\partial t_m}$ in~\eqref{eq:sylvester_equation_2} is characterized by the equation:
\begin{equation}\label{eq:sylvester_equation_12}
    \frac{\partial \mathbf{C}_{\mathtt{T}}^{1/2}}{\partial t_m} \mathbf{C}_{\mathtt{T}}^{\frac12}
    + \mathbf{C}_{\mathtt{T}}^{\frac12} \frac{\partial \mathbf{C}_{\mathtt{T}}^{1/2}}{\partial t_m}
    = \frac{\partial\mathbf{C}_{\mathtt{T}}}{\partial t_m}.
\end{equation}
Although directly computing the derivatives of matrix inverse square roots is challenging, the required derivatives can be obtained by solving~\eqref{eq:sylvester_equation_1},~\eqref{eq:sylvester_equation_2}, and~\eqref{eq:sylvester_equation_12}. These equations all take the standard form of Sylvester equations~\cite{higham2008functions} and can therefore be efficiently solved using dedicated solvers, such as the MATLAB function \texttt{sylvester}. In this way, computing the derivatives of the inverse square roots of the MC matrices is reduced to solving Sylvester equations. In conclusion, the first- and second-order partial derivatives of the objective function, $h_{\mathtt{T}}^{\prime}(t_m)$ and $h_{\mathtt{T}}^{\prime\prime}(t_m)$, are obtained by first substituting the solutions of~\eqref{eq:sylvester_equation_1},~\eqref{eq:sylvester_equation_2}, and~\eqref{eq:sylvester_equation_12} into~\eqref{eq:partial_h_t_m} and~\eqref{eq:partial2_h_t_m}, and then substituting the results into~\eqref{eq:h_prime_t_m} and~\eqref{eq:h_prime2_t_m}. With these derivatives, the local approximation $w_{\mathtt{T}}(t_m)$ is constructed according to~\eqref{eq:w_t_m}. Therefore, subproblem~\eqref{eq:t_m_problem} can be readily solved using the TRM. The procedure for updating $t_m$ is summarized in Algorithm~\ref{alg:t_m}.

\begin{algorithm}
    \caption{TRM-based algorithm for updating $t_m$.}\label{alg:t_m}
    \begin{algorithmic}[1]
        \Require $\{t_m\}_{m=1}^{M}$, $k$, $\sigma^2$, $\mathbf{\Sigma}$, $L$, $\{\theta_l^{\mathtt{T}}\}_{l=1}^{L}$, $D_\mathtt{T}$, $d_{\min}$, $\overline{\mathbf{Q}}$, $\overline{\mathbf{F}}$, $\overline{\mathbf{C}}_{\mathtt{R}}$
        \State Initialize $\mathcal{T}_m$ via~\eqref{eq:t_m_feasible_set}.
        \Repeat
            \State Calculate $\frac{\partial \mathbf{C}_{\mathtt{T}}^{-\frac12}}{\partial t_m}$ and $\frac{\partial^2 \mathbf{C}_{\mathtt{T}}^{-\frac12}}{\partial t_m^2}$ via~\eqref{eq:sylvester_equation_1}--\eqref{eq:sylvester_equation_12}.
            \State Calculate $\frac{\partial \mathbf{H}}{\partial t_m}$ and $\frac{\partial^2 \mathbf{H}}{\partial t_m^2}$ via~\eqref{eq:partial_h_t_m} and~\eqref{eq:partial2_h_t_m}.
            \State Calculate $\frac{\partial h_{\mathtt{T}}}{\partial t_m}$ and $\frac{\partial^2 h_{\mathtt{T}}}{\partial t_m^2}$ via~\eqref{eq:h_prime_t_m} and~\eqref{eq:h_prime2_t_m}.
            \State Find the local optimum $t_m^\star$ via~\eqref{eq:t_m_star}.
            \State Calculate $\varrho$ via~\eqref{eq:rho}.
            \If{$\varrho > \varrho_2$}
                \State $t_m \gets t_m^\star$.
                \If{$t_m$ is on the boundary of $\mathcal{T}_m$}
                    \State $A \gets \nu_1 A$.
                \EndIf
            \ElsIf{$\varrho > \varrho_1$}
                \State $t_m \gets t_m^\star$.
            \Else
                \State $A \gets A/\nu_2$.
            \EndIf
        \Until{the objective value $h_{\mathtt{T}}(t_m)$ converges.}
        \Ensure $t_m$.
    \end{algorithmic}
\end{algorithm}

\begin{remark}
Although only isotropic antenna elements are considered in this paper, the proposed TRM-based algorithm can be readily extended to more practical antenna types, such as dipoles. Changing the antenna model only affects the coupling matrix $\mathbf{C}_{\mathtt{T}}$. One simply needs to rederive~\eqref{eq:partial_c_t} and~\eqref{eq:partial2_c_t}, which involve only \textbf{element-wise} derivatives.
\end{remark}

\subsection{Update of $r_n$}\label{subsec:r_n}
In this step, we aim to optimize the $n$-th receive MA position $r_n$ while keeping the transmit covariance matrix $\mathbf{Q}$, the transmit MA positions $\mathbf{t}$, and the other receive MA positions $r_{n^\prime}\ (n^\prime \neq n)$ fixed. The optimization problem~\eqref{eq:se_max_problem} can be reformulated as
\begin{equation}\label{eq:r_n_problem}
    \setlength{\abovedisplayskip}{8pt}
    \setlength{\belowdisplayskip}{8pt}
    \max_{r_n}\ h_\mathtt{R}(r_n), \quad \text{s.t.}\ \overline{r}_{n-1}+d_{\min}\leq r_n \leq \overline{r}_{n+1}-d_{\min},
\end{equation}
where we define $\overline{r}_0 = -d_{\min}$ and $\overline{r}_{N+1} = D_{\mathtt{R}}+d_{\min}$. The objective function $h_{\mathtt{R}}(r_n)$ is defined as
\begin{equation*}
    h_{\mathtt{R}}(r_n) = \log \det\left(\mathbf{I}_M + \tfrac{1}{\sigma^2}
    \mathbf{H}^\hermconj(r_n) \overline{\mathbf{S}} \mathbf{H}(r_n)\right).
\end{equation*}
Here, $\overline{\mathbf{S}} = \overline{\mathbf{U}}\mathrm{diag}(P_1, P_2, \dots, P_\Gamma)\overline{\mathbf{U}}^\hermconj$.  
The channel reciprocity property in MIMO systems~\cite[Eq.~(21)]{maMIMOCapacityCharacterization2024} ensures that the objective $h_{\mathtt{R}}(r_n)$ equals the capacity $C$ after each update of $\mathbf{Q}$. The update of $r_n$ can be obtained using a similar TRM-based procedure as detailed in Section~\ref{subsec:t_m}. The only differences lie in the calculation of partial derivatives. The first- and second-order derivatives of $h_\mathtt{R}(r_n)$ can be obtained by interchanging $\mathbf{H}$ and $\mathbf{H}^\hermconj$ and substituting $\overline{\mathbf{Q}}$ with $\overline{\mathbf{S}}$ in~\eqref{eq:h_prime_t_m} and~\eqref{eq:h_prime2_t_m}, respectively. The first- and second-order derivatives of $\mathbf{H}$ w.r.t. $r_n$ are given by

\begin{small}
    \vspace{-0.3cm}
    \begin{align}
        \frac{\partial \mathbf{H}}{\partial r_n} &= \Big[\frac{\partial \mathbf{C}_{\mathtt{R}}^{-1/2}}{\partial r_n} \mathbf{F}^\hermconj + \mathbf{C}_{\mathtt{R}}^{-1/2} \frac{\partial \mathbf{F}^\hermconj}{\partial r_n}\Big] \mathbf{\Sigma} \overline{\mathbf{G}}\overline{\mathbf{C}}_{\mathtt{T}}^{-1/2}, \label{eq:partial_h_r_n}\\
        \frac{\partial^2 \mathbf{H}}{\partial r_n^2} &= \Big[\frac{\partial^2 \mathbf{C}_{\mathtt{R}}^{-1/2}}{\partial r_n^2} \mathbf{F}^\hermconj + 2\frac{\partial \mathbf{C}_{\mathtt{R}}^{-1/2}}{\partial r_n} \frac{\partial \mathbf{F}^\hermconj}{\partial r_n} + \mathbf{C}_{\mathtt{R}}^{-\frac12} \frac{\partial^2 \mathbf{F}^\hermconj}{\partial r_n^2}\Big] \mathbf{\Sigma} \overline{\mathbf{G}}\overline{\mathbf{C}}_{\mathtt{T}}^{-\frac12}, \label{eq:partial2_h_r_n}
    \end{align}
\end{small}%
respectively. The remaining derivations follow those in~\eqref{eq:partial_g_t_m}--\eqref{eq:sylvester_equation_12}, by replacing $t_m$ with $r_n$, $\theta_l^{\mathtt{T}}$ with $\theta_l^{\mathtt{R}}$, $\mathbf{G}$ with $\mathbf{F}$, and the subscript (or superscript) $\mathtt{T}$ with $\mathtt{R}$. The overall procedure for solving the optimization problem~\eqref{eq:se_max_problem} is summarized in Algorithm~\ref{alg:bca}.

\begin{small}
\begin{algorithm}
    \caption{BCA-based algorithm for solving problem~\eqref{eq:se_max_problem}.}\label{alg:bca}
    \begin{algorithmic}[1]
        \Require $k$, $P_{\max}$, $\sigma^2$, $\mathbf{\Sigma}$, $L$, $\{\theta_l^{\mathtt{T}}\}_{l=1}^{L}$, $\{\theta_l^{\mathtt{R}}\}_{l=1}^{L}$, $D_\mathtt{T}$, $D_\mathtt{R}$, $d_{\min}$.
        \State Initialize $\mathbf{t}$ and $\mathbf{r}$ with feasible values.
        \Repeat
            \State Update $\mathbf{Q}$ based on~\eqref{eq:q_opt}.
            \State Update each $t_m$ using the TRM in Algorithm~\ref{alg:t_m}.
            \State Update each $r_n$ using the TRM in Section~\ref{subsec:r_n}.
        \Until{the objective value $C$ converges.}
        \Ensure $\mathbf{Q}$, $\mathbf{t}$, $\mathbf{r}$.
    \end{algorithmic}
\end{algorithm}
\end{small}%

\subsection{Complexity and Convergence Analyses}\label{subsec:nb_analysis}
Algorithm~\ref{alg:bca} involves a two-loop optimization procedure, where the outer loop updates the transmit covariance matrix $\mathbf{Q}$ and the inner loops update the MA positions $t_m$ and $r_n$, respectively. The overall time complexity of Algorithm~\ref{alg:bca} is dominated by the TRM updates for $\mathbf{t}$ and $\mathbf{r}$. We therefore focus on one TRM iteration for updating a single MA position. Solving the Sylvester equations~\eqref{eq:sylvester_equation_1},~\eqref{eq:sylvester_equation_2}, and~\eqref{eq:sylvester_equation_12} each requires $\mathcal{O}(M^3)$ operations. Computing the first- and second-order partial derivatives of the channel matrix $\mathbf{H}$ w.r.t. $t_m$ via~\eqref{eq:partial_h_t_m} and~\eqref{eq:partial2_h_t_m} both incurs a time complexity of $\mathcal{O}(L M^2)$. Evaluating the first- and second-order derivatives of the objective function $h_{\mathtt{T}}(t_m)$ via~\eqref{eq:h_prime_t_m} and~\eqref{eq:h_prime2_t_m} each requires $\mathcal{O}(M^3)$ operations. Therefore, updating a single transmit MA position has a time complexity of $\mathcal{O}(M^3 + L M^2)$. Since there are $M$ transmit MA positions and $N$ receive MA positions, the total time complexity of Algorithm~\ref{alg:bca} is $\mathcal{O}(T_{\rm BCA} T_{\rm TRM} (M^3 (M+L) + N^3 (N+L)))$, where $T_{\rm BCA}$ and $T_{\rm TRM}$ denote the numbers of iterations of the outer BCA loop and the inner TRM loop, respectively.

We next analyze the convergence of Algorithm~\ref{alg:bca}. The update of the covariance matrix $\mathbf{Q}$ leads to a monotonic increase in capacity due to the concavity of $C$ w.r.t. $\mathbf{Q}$. The updates of the antenna positions $t_m$ and $r_n$ also lead to a monotonic increase in $C$, since any candidate update that reduces the capacity yields a negative $\varrho$ in~\eqref{eq:rho} and is rejected by the TRM. Moreover, the capacity $C$ is inherently upper-bounded by definition. The convergence of Algorithm~\ref{alg:bca} is thus guaranteed, since the capacity $C$ is non-decreasing and upper-bounded. The convergence behavior will be validated numerically in Fig.~\ref{fig:nb_conv}.

\section{MC-Aware Sum-Rate Maximization for Wideband Systems}\label{sec:capacity_max_wb}
We have discussed MC-aware capacity maximization for narrowband communication systems in Section~\ref{sec:capacity_max_nb}. However, the flat fading assumption does not always hold in practice. In wideband systems, frequency-selective fading leads to different optimal MA positions across subcarriers, thereby rendering the narrowband optimization algorithm inapplicable. In this section, we discuss the MC-aware sum-rate maximization problem for wideband systems and extend the BCA-based optimization framework and TRM-based algorithm to such systems.

\subsection{System Model}
When the flat fading assumption holds, the channel delay spread is negligible, and the channel matrix remains invariant in both the time and frequency domains, as shown in~\eqref{eq:system_model_mc}. When this assumption becomes invalid, the channel exhibits frequency selectivity, where the time-domain channel matrix depends on the delay and the frequency-domain channel matrix becomes frequency-dependent.

We employ an OFDM waveform to mitigate frequency-selective fading. The OFDM waveform consists of $S$ subcarriers with subcarrier spacing $\Delta$ and employs a pulse-shaping filter $f(t)$ that is non-zero only over the interval $[-1,1]$~\cite{irshadOptimizingMovableAntennas2025}. Suppose the channel consists of $L$ propagation paths with delays $\tau_l,\ l=1,2,\dots,L$. The maximum delay-tap index $T$ is given by $T = \lceil S\Delta(\tau_{\max} - \tau_{\min}) \rceil$, where $\tau_{\max} \triangleq \max\{\tau_l\}_{l=1}^L$ and $\tau_{\min} \triangleq \min\{\tau_l\}_{l=1}^L$ denote the maximum and minimum path delays, respectively. Under OFDM, the time-domain PRM $\mathbf{\Sigma}$ becomes delay-dependent:
\begin{equation*}
    \setlength{\abovedisplayskip}{8pt}
    \setlength{\belowdisplayskip}{8pt}
    \mathbf{\Sigma}_{\mathtt{td}}[\tau] = \mathrm{diag}\{b_1^{\mathtt{td}}[\tau], b_2^{\mathtt{td}}[\tau], \dots, b_L^{\mathtt{td}}[\tau]\}, \quad \tau = 0,1,\dots,T,
\end{equation*}
where $b_l^{\mathtt{td}}[\tau]$ denotes the time-domain complex gain of the $l$-th path at delay index $\tau$. The expression for $b_l^{\mathtt{td}}[\tau]$ is given by
\begin{multline}\label{eq:ofdm_prm_td}
    b_l^{\mathtt{td}}[\tau] = \alpha_l \mathrm{e}^{-jk(\tau_l - \tau_{\min})c} f(\tau - S\Delta(\tau_l - \tau_{\min})),\\ \tau = 0,1,\dots,T,
\end{multline}
where $\alpha_l$ and $c$ denote the complex gain of the $l$-th path and the speed of light, respectively. Similar to the narrowband case, the FRMs and MC matrices are still defined in~\eqref{eq:frm_def} and~\eqref{eq:mc_matrix_def}, respectively. Accordingly, the time-domain channel matrix is given by
\begin{equation}\label{eq:channel_matrix_wb_td}
    \mathbf{H}_{\mathtt{td}}(\mathbf{t}, \mathbf{r}, \tau) = \mathbf{C}_{\mathtt{R}}^{-\frac12}(\mathbf{r}) \mathbf{F}^\hermconj(\mathbf{r}) \mathbf{\Sigma}_{\mathtt{td}}[\tau] \mathbf{G}(\mathbf{t}) \mathbf{C}_{\mathtt{T}}^{-\frac12}(\mathbf{t}).
\end{equation}
To obtain the frequency-domain channel matrix, we apply the discrete Fourier transform (DFT) to $b_l^{\mathtt{td}}[\tau]$:
\begin{equation}\label{eq:ofdm_prm}
    b_l[\nu] = \sum_{\tau = 0}^{T} b_l^{\mathtt{td}}[\tau] \mathrm{e}^{-j2\pi\nu\tau/S}, \qquad \nu = 0,1,\dots,S-1,
\end{equation}
where $\nu$ denotes the subcarrier index. The resulting frequency-domain channel matrix is given by
\begin{equation}\label{eq:ofdm_channel_model}
    \mathbf{H}_\nu(\mathbf{t}, \mathbf{r}) = \mathbf{C}_{\mathtt{R}}^{-\frac12}(\mathbf{r}) \mathbf{F}^\hermconj(\mathbf{r}) \mathbf{\Sigma}_\nu \mathbf{G}(\mathbf{t}) \mathbf{C}_{\mathtt{T}}^{-\frac12}(\mathbf{t}),
\end{equation}
where $\mathbf{\Sigma}_\nu = \mathrm{diag}\{b_1[\nu], b_2[\nu], \dots, b_L[\nu]\}$ is the frequency-domain PRM.

The OFDM waveform enables parallel transmission across the $S$ subcarriers. The received signal at the $\nu$-th subcarrier is given by
\begin{equation}\label{eq:ofdm_signal_model}
    \mathbf{y}_\nu = \mathbf{H}_\nu(\mathbf{t}, \mathbf{r}) \mathbf{x}_\nu + \mathbf{n}_\nu, \qquad \nu = 0,1,\dots,S-1,
\end{equation}
where $\mathbf{x}_\nu$, $\mathbf{y}_\nu$, and $\mathbf{n}_\nu \sim \mathcal{CN}(\mathbf{0},\sigma_{\mathtt{w}}^2\mathbf{I}_N)$ denote the transmitted signal, received signal, and AWGN at the $\nu$-th subcarrier, respectively. Here, $\sigma_{\mathtt{w}}^2$ denotes the wideband noise variance. The transmit covariance matrix on each subcarrier is defined as $\mathbf{Q}_\nu \triangleq \E[\mathbf{x}_\nu \mathbf{x}_\nu^\hermconj] \in \mathbb{C}^{M\times M}$. The models in~\eqref{eq:ofdm_channel_model} and~\eqref{eq:ofdm_signal_model} provide a foundation for analyzing the impact of MC on the sum-rate performance of wideband MA-MIMO systems.

\subsection{Problem Formulation}
In OFDM systems, the capacity varies across subcarriers due to frequency-selective fading and independent transmit covariance matrices. Given the covariance matrix $\mathbf{Q}_\nu$, the capacity on the $\nu$-th subcarrier is given by
\begin{equation*}
    C_\nu(\mathbf{t}, \mathbf{r}, \mathbf{Q}_\nu) = \log \det\left(\mathbf{I}_N + \tfrac{1}{\sigma_{\mathtt{w}}^2}
    \mathbf{H}_\nu(\mathbf{t},\mathbf{r}) \mathbf{Q}_\nu \mathbf{H}_\nu^\hermconj(\mathbf{t},\mathbf{r})\right).
\end{equation*}
To eliminate inter-symbol interference (ISI) and inter-carrier interference (ICI), a cyclic prefix (CP) is employed. The CP length is typically chosen to exceed the maximum channel delay spread. Without loss of generality, we assume that the CP length $S_\mathrm{CP}$ equals $T$ throughout this paper. The overall OFDM performance is characterized by the sum of the subcarrier capacities. However, the CP introduces transmission overhead, reducing the effective sum-rate by a factor of $\frac{S}{S+S_\mathrm{CP}}$. Accordingly, the effective sum-rate is given by
\begin{equation*}
    \mathrm{SR}(\mathbf{t}, \mathbf{r}) = \frac{S}{S+S_\mathrm{CP}} \max_{\substack{\{\mathbf{Q}_\nu\}_{\nu=0}^{S-1}: \mathbf{Q}_\nu \succeq \mathbf{0} \\ \sum\limits_{\nu=0}^{S-1}\trace(\mathbf{Q}_\nu)\leq P_{\max}}} \sum_{\nu = 0}^{S-1} C_\nu(\mathbf{t}, \mathbf{r}, \mathbf{Q}_\nu).
\end{equation*}
Similar to the narrowband capacity $C(\mathbf{t}, \mathbf{r})$, the sum-rate $\mathrm{SR}(\mathbf{t}, \mathbf{r})$ remains a function of the transmit and receive MA positions $\mathbf{t}$ and $\mathbf{r}$. The resulting sum-rate maximization problem is formulated as\footnote{Note that the narrowband system investigated in Section~\ref{sec:capacity_max_nb} can be mathematically viewed as a special case of this wideband formulation by setting $S=1$. We dedicate a separate section to the narrowband case to isolate the fundamental mathematical challenge of MC matrix optimization from the difficulty of balancing varying optimal antenna positions across multiple subcarriers.}

\begin{small}
    \vspace{-0.3cm}
    \begin{subequations}\label{eq:ofdm_sr_max_problem}
        \begin{align}
            \max_{\mathbf{Q}_\nu,\mathbf{t},\mathbf{r}} \quad &
            \frac{S}{S+S_\mathrm{CP}}\sum_{\nu=0}^{S-1} \log \det\!\left(\mathbf{I}_N + \tfrac{1}{\sigma_{\mathtt{w}}^2}
            \mathbf{H}_\nu(\mathbf{t},\mathbf{r}) \mathbf{Q}_\nu \mathbf{H}_\nu^\hermconj(\mathbf{t},\mathbf{r})\right), \label{eq:ofdm_sr_max_obj}\\
            \text{s.t.} \quad &
            \mathbf{Q}_\nu \succeq \mathbf{0}, \quad \nu = 0,1,\dots,S-1, \label{eq:ofdm_sr_max_q_psd}\\
            &
            \sum_{\nu=0}^{S-1} \trace(\mathbf{Q}_\nu) \leq P_{\max}, \label{eq:ofdm_sr_max_power}\\
            &\eqref{eq:se_max_t_range}-\eqref{eq:se_max_r_spacing}, \nonumber
        \end{align}
    \end{subequations}
\end{small}%
where constraints~\eqref{eq:ofdm_sr_max_q_psd} and~\eqref{eq:ofdm_sr_max_power} ensure the positive semidefiniteness of $\mathbf{Q}_\nu$ and that the total transmitted power across all subcarriers does not exceed $P_{\max}$, respectively. To ensure a constant average transmit SNR as the number of subcarriers $S$ varies, we define $P_{\max} = \rho S \Delta$ with a slight abuse of notation, where $\rho$ denotes the transmit power density per MHz. The constraints on the MA positions are identical to those in the narrowband capacity maximization problem~\eqref{eq:se_max_problem}, given in~\eqref{eq:se_max_t_range}--\eqref{eq:se_max_r_spacing}.

The TRM-based algorithm proposed in Section~\ref{sec:capacity_max_nb} for solving the narrowband capacity maximization problem~\eqref{eq:se_max_problem} cannot be directly applied to the wideband sum-rate maximization problem~\eqref{eq:ofdm_sr_max_problem}. The narrowband TRM-based algorithm can only optimize the MA positions to maximize the capacity of a specific subcarrier. If $t_m$ is optimized to maximize the capacity of the $\nu$-th subcarrier, it is not guaranteed that the same $t_m$ also maximizes the capacities of the other subcarriers. Since different subcarriers share the same set of antenna configurations, the MA position optimization cannot be decomposed into independent per-subcarrier subproblems. Instead, the objective function~\eqref{eq:ofdm_sr_max_obj} requires finding a single set of transmit and receive MA positions that effectively balances the varying channel conditions of different subcarriers. In what follows, we extend the BCA-based framework and the TRM-based algorithm to wideband systems to meet the requirements.

\subsection{Update of $\mathbf{Q}_\nu$}
In this step, we optimize the transmit covariance matrices $\{\mathbf{Q}_\nu\}_{\nu=0}^{S-1}$ on each subcarrier while keeping the MA positions $\mathbf{t}$ and $\mathbf{r}$ fixed. If we omit constant factors, the optimization problem can be formulated as
\begin{subequations}\label{eq:ofdm_sr_max_q_problem}
    \begin{align}
        \max_{\mathbf{Q}_\nu} \quad & \sum_{\nu=0}^{S-1} \log \det\left(\mathbf{I}_N + \tfrac{1}{\sigma_{\mathtt{w}}^2} \overline{\mathbf{H}}_\nu \mathbf{Q}_\nu \overline{\mathbf{H}}_\nu^\hermconj\right), \\
        \text{s.t.} \quad & \mathbf{Q}_\nu \succeq \mathbf{0}, \quad \nu = 0,1,\dots,S-1, \\
        & \sum_{\nu=0}^{S-1} \trace(\mathbf{Q}_\nu) \leq P_{\max},
    \end{align}
\end{subequations}
where we define $\overline{\mathbf{H}}_\nu = \mathbf{H}_{\nu}(\overline{\mathbf{t}}, \overline{\mathbf{r}})$ for simplicity. The subproblem~\eqref{eq:ofdm_sr_max_q_problem} is concave and can be solved using the water-filling algorithm. Let the SVD of $\overline{\mathbf{H}}_\nu$ be expressed as $\overline{\mathbf{H}}_\nu = \overline{\mathbf{U}}_\nu \overline{\mathbf{\Lambda}}_\nu \overline{\mathbf{V}}_\nu^\hermconj$, where $\overline{\mathbf{\Lambda}}_\nu = \mathrm{diag}\{\lambda_1^{\nu}, \lambda_2^{\nu}, \dots, \lambda_{\Gamma_\nu}^{\nu}\} \in \mathbb{R}^{\Gamma_\nu \times \Gamma_\nu}$, $\overline{\mathbf{U}}_\nu\in\mathbb{C}^{N\times \Gamma_\nu}$, $\overline{\mathbf{V}}_\nu\in\mathbb{C}^{M\times \Gamma_\nu}$, and $\Gamma_\nu$ denotes the rank of $\overline{\mathbf{H}}_\nu$. According to the water-filling principle, the transmit power allocated to the $\gamma$-th eigenchannel of the $\nu$-th subcarrier is given by
\begin{equation*}
    P_\gamma^\nu = \max\left(0, \mu - \tfrac{\sigma_{\mathtt{w}}^2}{(\lambda_\gamma^{\nu})^2}\right),
\end{equation*}
where $\gamma = 1,2,\dots,\Gamma_\nu$, $\nu = 0,1,\dots,S-1$, and $\mu$ is the water level. The water level $\mu$ is chosen by bisection search to satisfy the total power constraint $\sum_{\nu=0}^{S-1} \sum_{\gamma=1}^{\Gamma_\nu} P_\gamma^\nu = P_{\max}$. With the water level $\mu$, the transmit covariance matrix of the $\nu$-th subcarrier is given by
\begin{equation}\label{eq:ofdm_q_opt}
    \mathbf{Q}_\nu = \overline{\mathbf{V}}_\nu \mathrm{diag}(P_1^\nu, P_2^\nu, \dots, P_{\Gamma_\nu}^\nu) \overline{\mathbf{V}}_\nu^\hermconj.
\end{equation}

\subsection{Update of $\mathbf{t}$ and $\mathbf{r}$}\label{subsec:ofdm_pos}
In this step, we discuss the update steps of the MA positions $\mathbf{t}$ and $\mathbf{r}$. We first optimize the $m$-th transmit MA position $t_m$ while keeping the remaining MA positions and all the transmit covariance matrices fixed. The resulting optimization problem is formulated as
\begin{subequations}\label{eq:ofdm_t_m_problem}
    \begin{align}
        \max_{t_m}\quad & h_{\mathtt{T}, \mathtt{w}}(t_m) \triangleq \sum_{\nu=0}^{S-1} h_{\nu, \mathtt{T}}(t_m),\label{eq:ofdm_t_m_obj}\\
        \text{s.t.}\quad &\overline{t}_{m-1}+d_{\min}\leq t_m \leq \overline{t}_{m+1}-d_{\min},
    \end{align}
\end{subequations}
where $h_{\nu, \mathtt{T}}(t_m)$ denotes the capacity of the $\nu$-th subcarrier with $t_m$ treated as the only optimization variable. The expression for $h_{\nu, \mathtt{T}}(t_m)$ is given by
\begin{equation*}
    h_{\nu, \mathtt{T}}(t_m) = \log \det\left(\mathbf{I}_N + \tfrac{1}{\sigma_{\mathtt{w}}^2} \mathbf{H}_\nu(t_m) \overline{\mathbf{Q}}_\nu \mathbf{H}_\nu^\hermconj(t_m)\right).
\end{equation*}

Similar to the narrowband scenario, the objective function $h_{\mathtt{T}, \mathtt{w}}(t_m)$ is highly non-concave. We extend the TRM-based algorithm to solve subproblem~\eqref{eq:ofdm_t_m_problem} in the wideband case. The key to extending the TRM-based algorithm lies in constructing an approximation function $w_{\mathtt{T}, \mathtt{w}}(t_m)$ that accounts for the capacities of all subcarriers. The approximation function $w_{\mathtt{T}, \mathtt{w}}(t_m)$ can be constructed by summing the per-subcarrier approximation functions $w_{\nu, \mathtt{T}}(t_m)$ as follows:

\begin{small}
    \vspace{-0.3cm}
    \begin{align}\label{eq:ofdm_w_t_m}
        w_{\mathtt{T}, \mathtt{w}}(t_m) &= \sum_{\nu=0}^{S-1} w_{\nu, \mathtt{T}}(t_m) \nonumber \\
        &= \sum_{\nu=0}^{S-1}\left[\tfrac12 h_{\nu, \mathtt{T}}^{\prime\prime}(\overline{t}_m) (t_m - \overline{t}_m)^2 + h_{\nu, \mathtt{T}}^{\prime}(\overline{t}_m) (t_m - \overline{t}_m)\right],
    \end{align}
\end{small}%
where the constant term is omitted for simplicity. The approximation function $w_{\mathtt{T}, \mathtt{w}}(t_m)$ requires only the first- and second-order derivatives of $h_{\nu, \mathtt{T}}(t_m)$, which are given by
\begin{align}
    &h_{\nu, \mathtt{T}}^\prime(t_m) = \tfrac2{\sigma_{\mathtt{w}}^2} \Re\left\{\trace\left[\mathbf{\Phi}_\nu \tfrac{\partial \mathbf{H}_\nu}{\partial t_m} \overline{\mathbf{Q}}_\nu \mathbf{H}_\nu^\hermconj\right] \right\},\label{eq:ofdm_h_t_m_prime}\\
    &h_{\nu, \mathtt{T}}^{\prime\prime}(t_m) \nonumber \\
    =&\tfrac2{\sigma_{\mathtt{w}}^2}\Re\left\{\trace\left[\mathbf{\Phi}_\nu\left(\tfrac{\partial^2 \mathbf{H}_\nu}{\partial t_m^2} \overline{\mathbf{Q}}_\nu \mathbf{H}_\nu^\hermconj + \tfrac{\partial \mathbf{H}_\nu}{\partial t_m} \overline{\mathbf{Q}}_\nu \tfrac{\partial \mathbf{H}_\nu^\hermconj}{\partial t_m}\right)\right. \right.\nonumber \\
    &\hspace{3em}-\tfrac{1}{\sigma_{\mathtt{w}}^2}\mathbf{\Phi}_\nu\tfrac{\partial \mathbf{H}_\nu}{\partial t_m} \overline{\mathbf{Q}}_\nu\mathbf{H}_\nu^\hermconj \mathbf{\Phi}_\nu\tfrac{\partial \mathbf{H}_\nu}{\partial t_m}\overline{\mathbf{Q}}_\nu\mathbf{H}_\nu^\hermconj \nonumber \\
    &\hspace{3em}-\left.\left.\tfrac{1}{\sigma_{\mathtt{w}}^2} \mathbf{\Phi}_\nu\tfrac{\partial \mathbf{H}_\nu}{\partial t_m} \overline{\mathbf{Q}}_\nu\mathbf{H}_\nu^\hermconj \mathbf{\Phi}_\nu\mathbf{H}_\nu\overline{\mathbf{Q}}_\nu\tfrac{\partial \mathbf{H}_\nu^\hermconj}{\partial t_m}\right]\right\},\label{eq:ofdm_h_t_m_prime2}
\end{align}
respectively. Here, we define $\mathbf{\Phi}_\nu \triangleq (\mathbf{I}_N + \frac1{\sigma_{\mathtt{w}}^2}\mathbf{H}_\nu\overline{\mathbf{Q}}_\nu\mathbf{H}_\nu^\hermconj)^{-1}$. The first- and second-order partial derivatives of $\mathbf{H}_\nu$ w.r.t. $t_m$ follow the same form as in the narrowband system, given in~\eqref{eq:partial_h_t_m} and~\eqref{eq:partial2_h_t_m}, with $\mathbf{\Sigma}_\nu$ replacing $\mathbf{\Sigma}$. The derivatives of the FRM $\mathbf{G}$ and the MC matrix $\mathbf{C}_{\mathtt{T}}$ remain unchanged and are described in~\eqref{eq:partial_g_t_m}--\eqref{eq:sylvester_equation_12}.

Next, we optimize the $n$-th receive MA position $r_n$ while keeping the remaining MA positions and all the transmit covariance matrices fixed. The resulting optimization problem is formulated as
\begin{subequations}\label{eq:ofdm_r_n_problem}
    \begin{align}
        \max_{r_n}\quad & h_{\mathtt{R}, \mathtt{w}}(r_n) \triangleq \sum_{\nu=0}^{S-1} h_{\nu, \mathtt{R}}(r_n),\label{eq:ofdm_r_n_obj}\\
        \text{s.t.}\quad &\overline{r}_{n-1}+d_{\min}\leq r_n \leq \overline{r}_{n+1}-d_{\min},
    \end{align}
\end{subequations}
where $h_{\nu, \mathtt{R}}(r_n)$ denotes the capacity of the $\nu$-th subcarrier with $r_n$ treated as the sole optimization variable. The expression for $h_{\nu, \mathtt{R}}(r_n)$ is given by
\begin{equation*}
    h_{\nu, \mathtt{R}}(r_n) = \log \det\left(\mathbf{I}_M + \tfrac{1}{\sigma_{\mathtt{w}}^2}
    \mathbf{H}_\nu^\hermconj(r_n) \overline{\mathbf{S}}_\nu \mathbf{H}_\nu(r_n)\right).
\end{equation*}
Here, $\overline{\mathbf{S}}_\nu = \overline{\mathbf{U}}_\nu\mathrm{diag}(P_1^\nu, P_2^\nu, \dots, P_{\Gamma_\nu}^\nu)\overline{\mathbf{U}}_\nu^\hermconj$. To extend the TRM-based algorithm to subproblem~\eqref{eq:ofdm_r_n_problem}, we construct an approximation function $w_{\mathtt{R}, \mathtt{w}}(r_n)$ similar to~\eqref{eq:ofdm_w_t_m}, which only requires the first- and second-order derivatives of $h_{\nu, \mathtt{R}}(r_n)$. These derivatives follow the same structure as those in~\eqref{eq:ofdm_h_t_m_prime} and~\eqref{eq:ofdm_h_t_m_prime2}, respectively, with $\mathbf{H}_\nu$ and $\mathbf{H}_\nu^\hermconj$ interchanged, and $\overline{\mathbf{Q}}_\nu$ replaced by $\overline{\mathbf{S}}_\nu$. The first- and second-order partial derivatives of $\mathbf{H}_\nu$ w.r.t. $r_n$ are obtained by replacing $\mathbf{\Sigma}$ with $\mathbf{\Sigma}_\nu$ in~\eqref{eq:partial_h_r_n} and~\eqref{eq:partial2_h_r_n}, respectively. The overall procedure for solving problem~\eqref{eq:ofdm_sr_max_problem} is summarized in Algorithm~\ref{alg:ofdm_bca}.

\begin{algorithm}
    \caption{BCA-based algorithm for solving problem~\eqref{eq:ofdm_sr_max_problem}.}\label{alg:ofdm_bca}
    \begin{algorithmic}[1]
        \Require $k$, $S$, $S_{\rm CP}$, $\Delta$, $\rho$, $\sigma_{\mathtt{w}}^2$, $\mathbf{\Sigma}_\nu$, $L$, $\{\theta_l^{\mathtt{T}}\}_{l=1}^{L}$, $\{\theta_l^{\mathtt{R}}\}_{l=1}^{L}$, $D_\mathtt{T}$, $D_\mathtt{R}$, $d_{\min}$.
        \State Initialize $\mathbf{t}$ and $\mathbf{r}$ with feasible values.
        \Repeat
            \State Update each $\mathbf{Q}_\nu$ based on~\eqref{eq:ofdm_q_opt}.
            \State Update each $t_m$ using the TRM in Section~\ref{subsec:ofdm_pos}.
            \State Update each $r_n$ using the TRM in Section~\ref{subsec:ofdm_pos}.
        \Until{the objective value $\mathrm{SR}$ converges.}
        \Ensure $\mathbf{Q}_\nu$, $\mathbf{t}$, $\mathbf{r}$.
    \end{algorithmic}
\end{algorithm}

\begin{remark}\label{rem:capacity_gain}
    The potential performance gains of leveraging MC in both wideband and narrowband MA-MIMO systems are twofold. First, in MA systems, MC-induced superdirectivity can result in higher effective transmit and receive power~\cite{morrisSuperdirectivityMIMOSystems2005}. When MAs move off the $\lambda/2$ grid, the eigenvalues of the MC matrices $\mathbf{C}_{\mathtt{T}}(\mathbf{t})$ and $\mathbf{C}_{\mathtt{R}}(\mathbf{r})$ are no longer uniformly equal to $1$. Smaller eigenvalues of these matrices yield larger eigenvalues of $\mathbf{C}_{\mathtt{T}}^{-1/2}(\mathbf{t})$ and $\mathbf{C}_{\mathtt{R}}^{-1/2}(\mathbf{r})$, which in turn lead to higher transmit and receive power densities in the directions of $\{\theta_l^{\mathtt{T}}\}_{l=1}^{L}$ and $\{\theta_l^{\mathtt{R}}\}_{l=1}^{L}$, respectively. Second, unlike conventional MIMO systems with MC, MA makes the MC matrices $\mathbf{C}_{\mathtt{T}}$ and $\mathbf{C}_{\mathtt{R}}$ designable. Position optimization shapes the MC matrices to align with the eigenmodes of the wireless channel, resulting in a more uniform eigenvalue spectrum of the SNR matrix $\mathbf{H}\mathbf{Q}\mathbf{H}^\hermconj/\sigma^2$ and higher performance. 
\end{remark}

\subsection{Complexity and Convergence Analyses}
Similar to Algorithm~\ref{alg:bca} discussed in Section~\ref{subsec:nb_analysis}, the time complexity of Algorithm~\ref{alg:ofdm_bca} is also dominated by the TRM updates for $\mathbf{t}$ and $\mathbf{r}$. The only difference is that the calculation of the derivatives of the objective function $h_{\mathtt{T}, \mathtt{w}}(t_m)$ and $h_{\mathtt{R}, \mathtt{w}}(r_n)$ involves the sum of the per-subcarrier derivatives, which leads to a time complexity of $\mathcal{O}(S M^3 + S L M^2)$. In comparison, the time complexity of solving the Sylvester equations remains unchanged and can be neglected since it is invariant to the number of subcarriers $S$. Therefore, the time complexities of updating a single transmit and a single receive MA position are $\mathcal{O}(S M^3 + S L M^2)$ and $\mathcal{O}(S N^3 + S L N^2)$, respectively. The total time complexity of Algorithm~\ref{alg:ofdm_bca} is $\mathcal{O}(T_{\rm BCA} T_{\rm TRM}S (M^3(M+L) + N^3(N+L)))$. The convergence of Algorithm~\ref{alg:ofdm_bca} is guaranteed for the same reasons as Algorithm~\ref{alg:bca}. The convergence behaviors will be further validated in Fig.~\ref{fig:ofdm_conv}.

\section{Simulation Results}\label{sec:sim}
In this section, we evaluate the capacity improvements achieved by leveraging MC in MA-MIMO systems in both narrowband and wideband scenarios. We compare the capacity of the proposed algorithm, denoted by C-MA, with several baseline schemes.
\begin{itemize}
    \item \textbf{Uniform linear array (ULA)}: Fixed-position antennas (FPAs) with antenna spacing of $\lambda/2$. In this case, the MC matrix equals the identity matrix, and MC effects do not exist.
    \item \textbf{Compact linear array (CLA)}: FPAs with antenna spacing of $0.2\lambda$. Since the antenna spacing equals $d_{\min}$, its MC is comparable to, if not stronger than, that of C-MA.
    \item \textbf{No MC-MA (NC-MA)}: No MC is considered in the optimization for MA systems. Unless otherwise specified, the minimum antenna separation is set to $d_{\min}=\lambda/2$ to suppress MC. For narrowband systems, we adopt the method in~\cite{maMIMOCapacityCharacterization2024}. For OFDM systems, we use Algorithm~\ref{alg:ofdm_bca} with the MC matrices set to identity matrices.
\end{itemize}

\subsection{Simulation Setup}
\begin{table}[!t]
    \centering
    \caption{Key Simulation Parameters~\cite{irshadOptimizingMovableAntennas2025,3gpp.25.996}}\label{tab:sim_params}
    \begin{small}
    \begin{tabular}{|l|l|}
        \hline
        \textbf{Parameter} & \textbf{Value} \\ \hline
        Number of channel realizations & $N_s = 1000$ \\
        Number of transmit MAs & $M = 8$ \\
        Number of receive MAs & $N = 8$ \\ 
        Carrier frequency & $f_c = 28$\,GHz \\ 
        Minimum Tx-Rx distance & $r_{\min} = 100$\,m \\
        Maximum Tx-Rx distance & $r_{\max} = 300$\,m \\
        Distance between the Tx and Rx & $r^2 \sim \mathcal{U}[r_{\min}^2, r_{\max}^2]$ \\
        Rician factor & $\kappa = 0$\,dB \\
        Number of clusters & $L_{\mathrm{clu}} = 3$ \\
        Number of sub-paths per cluster & $L_{\mathrm{sub}} = 8$ \\
        AoD/AoA of the LoS path & $\theta_{\mathtt{LoS}}^{\mathtt{T}}, \theta_{\mathtt{LoS}}^{\mathtt{R}}\sim\mathcal{U}[-5^\circ, 5^\circ]$ \\
        AoD/AoA of cluster $i$ w.r.t. LoS & $\vartheta_{i}^{\mathtt{T}}, \vartheta_{i}^{\mathtt{R}}\sim\mathcal{U}[-40^\circ, 40^\circ]$ \\
        AoD/AoA of sub-path $p$ in cluster $i$ & $\delta_{i,p}^{\mathtt{T}}, \delta_{i,p}^{\mathtt{R}}\sim\mathcal{U}[-5^\circ, 5^\circ]$ \\
        Minimum MA separation & $d_{\min} = 0.2\lambda$ \\
        Transmit MA movement range & $D_\mathtt{T} = 2(M-1)\lambda$ \\
        Receive MA movement range & $D_\mathtt{R} = 2(N-1)\lambda$ \\
        Narrowband transmit power budget & $P_{\max} = 30$\,dBm \\
        Narrowband noise power & $\sigma^2 = -80$\,dBm \\
        Wideband transmit power density & $\rho = 0$\,dBm/MHz \\ 
        Wideband noise PSD & $N_0 = -174$\,dBm/Hz \\
        Noise factor & $\mathrm{NF} = 5$\,dB \\
        Number of subcarriers & $S = 300$ \\
        Subcarrier spacing & $\Delta = 15$\,kHz \\
        TRM thresholds & $\varrho_1 = 0.25, \varrho_2 = 0.75$ \\
        TRM factors & $\nu_1 = 2, \nu_2 = 4$ \\
        \hline
    \end{tabular}
    \end{small}
\end{table}

Unless otherwise specified, the simulation parameters are listed in Table~\ref{tab:sim_params}. Motivated by the urban micro scenario in~\cite{3gpp.25.996}, we adopt a Rician channel model consisting of one LoS component and $L_{\rm clu}$ scattering clusters. Each cluster contains $L_{\rm sub}$ sub-paths. Let $\theta_{\mathtt{LoS}}^{\mathtt{T}}$ and $\theta_{\mathtt{LoS}}^{\mathtt{R}}$ denote the AoD and AoA of the LoS component, respectively. Let $\vartheta_{i}^{\mathtt{T}}$ and $\vartheta_{i}^{\mathtt{R}}$ denote the relative AoD and AoA of the $i$-th scattering cluster center w.r.t. the LoS component, and let $\delta_{i,p}^{\mathtt{T}}$ and $\delta_{i,p}^{\mathtt{R}}$ denote the relative AoD and AoA of the sub-path $p$ w.r.t. the $i$-th scattering cluster. Then, the number of multipath components is given by $L = 1+L_{\rm clu}\cdot L_{\rm sub}$. For the LoS component, the AoD and AoA satisfy $\theta_1^{\mathtt{T}} = \theta_{\mathtt{LoS}}^{\mathtt{T}}$ and $\theta_1^{\mathtt{R}} = \theta_{\mathtt{LoS}}^{\mathtt{R}}$. For the non-LoS (NLoS) components, the AoDs and AoAs are given by $\theta_{1+(i-1)L_{\rm sub}+p}^{\mathtt{T}} = \theta_{\mathtt{LoS}}^{\mathtt{T}} + \vartheta_{i}^{\mathtt{T}} + \delta_{i,p}^{\mathtt{T}}$ and $\theta_{1+(i-1)L_{\rm sub}+p}^{\mathtt{R}} = \theta_{\mathtt{LoS}}^{\mathtt{R}} + \vartheta_{i}^{\mathtt{R}} + \delta_{i,p}^{\mathtt{R}}$, respectively, where $1\leq i\leq L_{\rm clu}$ and $1\leq p\leq L_{\rm sub}$. The path loss is given by $\mathrm{PL} = 32.4 + 20\log_{10}(f_c / 1~\mathrm{GHz}) + 26\log_{10}(r / 1~\mathrm{m}) + \delta_{\rm SF}$~[dB], where $\delta_{\rm SF}\sim\mathcal{N}(0, 16)$ is the shadowing factor. The delay of each cluster is uniformly distributed between $1$ and $10$ times the LoS delay, and all sub-paths within the same cluster share the corresponding cluster delay. The complex gain of the LoS component follows $\alpha_{\rm LoS} \sim \mathcal{CN}\left(0, \frac{\kappa}{1+\kappa}\cdot \mathrm{PL}^{-1}\right)$. The power delay profile of the $i$-th scattering cluster is given by $q_i = 10^{-\frac{\tau_i}{1\ \mathrm{\mu s}} + \frac{z_i}{10}} / \sum_{l=1}^{L_{\rm clu}} 10^{-\frac{\tau_l}{1\ \mathrm{\mu s}} + \frac{z_l}{10}}$, where $z_i \sim \mathcal{N}(0, 9)$. The complex gain of each sub-path within the $i$-th scattering cluster follows $\alpha_{i, p} \sim \mathcal{CN} \left(0, \frac{1}{\kappa+1}\frac{\mathrm{PL}^{-1}}{L_{\rm clu}\cdot L_{\rm sub}}\right)$. We define $\alpha_1 = \alpha_{\rm LoS}$ for the LoS component, and $\alpha_{1+(i-1)L_{\rm sub}+p} = \alpha_{i, p}$ for the NLoS components, where $1\leq i\leq L_{\rm clu}$ and $1\leq p\leq L_{\rm sub}$. In narrowband systems, $b_l = \alpha_l$, whereas in wideband systems, the relationships between $b_l$ and $\alpha_l$ are given in~\eqref{eq:ofdm_prm_td} and~\eqref{eq:ofdm_prm}. The wideband noise power on the $\nu$-th subcarrier is given by $\sigma_{\mathtt{w}}^2 = N_0 \Delta \cdot 10^{\mathrm{NF}/10}$, and the pulse-shaping filter is designed as $f(t) = 1 - \vert t\vert$ for $t\in[-1, 1]$~\cite{irshadOptimizingMovableAntennas2025}.

\subsection{Narrowband System}
\begin{figure}[!t]
    \centering
    \includegraphics[width=0.45\textwidth]{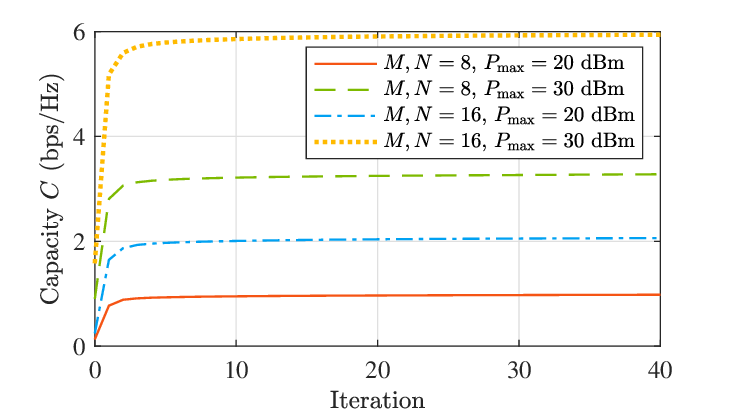}
    \caption{Convergence behaviors of Algorithm~\ref{alg:bca} in narrowband scenarios.}
    \label{fig:nb_conv}
\end{figure}

\begin{figure}[!t]
    \centering
    \includegraphics[width=0.45\textwidth]{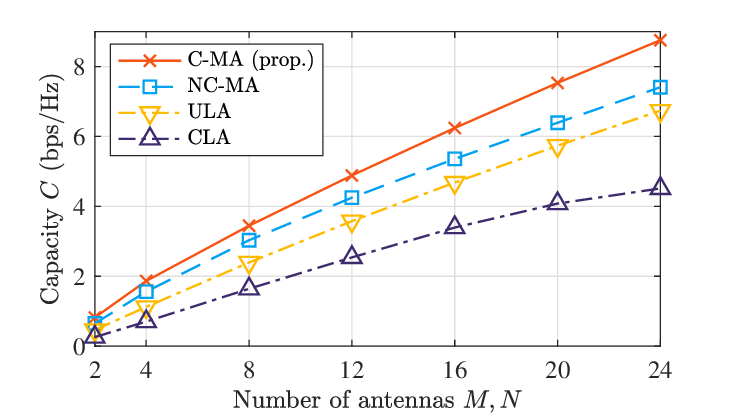}
    \caption{Impact of number of transmit MAs $M$ and receive MAs $N$ on the capacity $C$ of narrowband MC-aware MA systems ($M = N$).}
    \label{fig:nb_ntx}
\end{figure}

\begin{figure}[!t]
    \centering
    \includegraphics[width=0.45\textwidth]{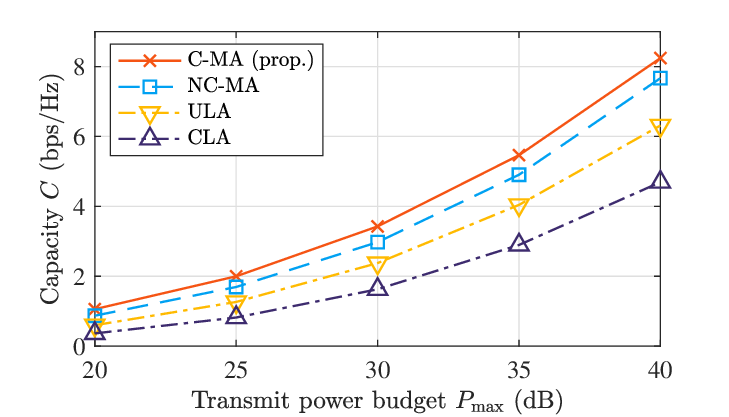}
    \caption{Impact of transmit power budget $P_{\max}$ on the capacity $C$ of narrowband MC-aware MA systems.}
    \label{fig:nb_power}
\end{figure}

Fig.~\ref{fig:nb_conv} shows the convergence behaviors of Algorithm~\ref{alg:bca} in narrowband scenarios. For different numbers of MAs and transmit power budgets, the capacity $C$ of Algorithm~\ref{alg:bca} increases monotonically with the number of BCA iterations. Regardless of the number of MAs and the transmit power budget, Algorithm~\ref{alg:bca} converges within around $20$ iterations. Fig.~\ref{fig:nb_ntx} illustrates the impact of the numbers of transmit MAs $M$ and receive MAs $N$ on the capacity $C$, where we set $M = N$ without loss of generality. C-MA achieves the highest capacity among all baseline schemes across different antenna configurations. When $M = N = 8$, the capacity of C-MA is approximately $13.7\,\%$ higher than that of NC-MA. When the number of MAs grows to $M = N = 24$, the capacity gain of C-MA over NC-MA increases to $18.1\,\%$. The capacity gain arises from superdirectivity, which enables beam gains exceeding the number of antennas~\cite{ivrlacCircuitTheoryCommunication2010}. Superdirectivity alone does not necessarily lead to improved performance. Although CLA also exhibits superdirectivity, its capacity is significantly lower than that of C-MA. This is because the small aperture of CLA limits its angular resolution, leading to inferior multiplexing gains~\cite{tse2005fundamentals}. Moreover, antenna movement allows the MC matrices to be adaptively shaped to better match the wireless channel and thereby achieve higher capacity. We next investigate the impact of the transmit power budget $P_{\max}$ on the capacity $C$ of MC-aware MA systems in Fig.~\ref{fig:nb_power}. For different transmit power budgets, C-MA outperforms all baselines. The capacity gain of C-MA over NC-MA is larger in low SNR regimes. When $P_{\max} = 20$\,dBm, the capacity of C-MA is around $21.5\,\%$ higher than that of NC-MA. When $P_{\max} = 40$\,dBm, the capacity gain of C-MA over NC-MA reduces to $7.49\,\%$. In low-SNR regimes, water-filling tends to allocate power only to the strongest eigenmodes. Therefore, higher directivity greatly enhances the strongest eigenmode, making superdirectivity more beneficial in this regime. In high-SNR regimes, by contrast, power allocation tends to be uniform, and the capacity gain provided by superdirectivity becomes less significant~\cite{zhang2025capacity}.

\begin{figure}[!t]
    \centering
    \includegraphics[width=0.45\textwidth]{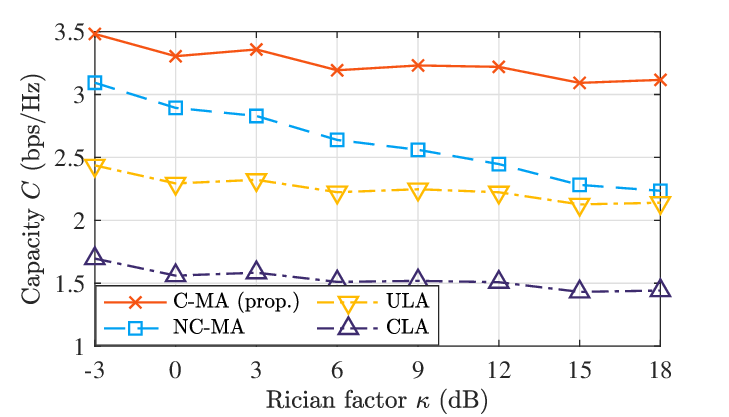}
    \caption{Impact of Rician factor $\kappa$ on the capacity $C$ of narrowband MC-aware MA systems.}
    \label{fig:nb_kappa}
\end{figure}

\begin{figure}[!t]
    \centering
    \includegraphics[width=0.45\textwidth]{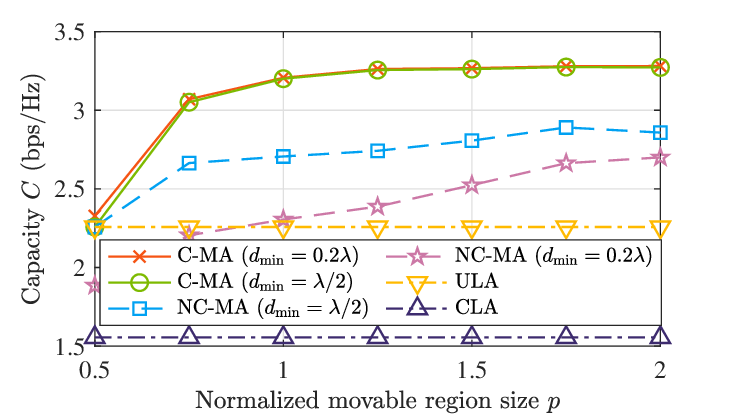}
    \caption{Impact of normalized movable range $p$ on the capacity $C$ of narrowband MC-aware MA systems.}
    \label{fig:nb_mr}
\end{figure}

Fig.~\ref{fig:nb_kappa} illustrates the impact of the Rician factor $\kappa$ on the capacity $C$ of MC-aware MA systems in narrowband scenarios. Larger values of $\kappa$ correspond to stronger LoS components, whereas smaller values of $\kappa$ correspond to stronger NLoS conditions. In both LoS-dominant and NLoS-dominant scenarios, C-MA achieves the highest capacity among all baselines. Moreover, the capacity gain of NC-MA over ULA becomes less pronounced with a stronger LoS component. This is because the SNR fluctuation induced by small-scale fading is weaker in strong LoS scenarios, and the SNR gain provided by antenna movement is therefore limited~\cite{zhuModelingPerformanceAnalysis2024}. However, in LoS-dominant scenarios, C-MA significantly outperforms NC-MA and ULA owing to the higher directivity enabled by superdirectivity. In Fig.~\ref{fig:nb_mr}, we depict the impact of the normalized movable range on the capacity $C$ of narrowband MC-aware MA systems, where $p = \frac{D_{\mathtt{T}}}{(M-1)\lambda} = \frac{D_{\mathtt{R}}}{(N-1)\lambda}$. The capacities of both C-MA and NC-MA increase with $p$ due to the enhanced antenna-movement DoFs. When $p = 0.5$, the capacity of NC-MA with minimum MA separation $d_{\min} = \lambda/2$ is identical to that of ULA, since NC-MA cannot move. By contrast, C-MA with $d_{\min} = 0.2\lambda$ still achieves a capacity gain of approximately $3.13\,\%$ over ULA. This observation indicates that under a limited movable range, the increased antenna-movement DoFs can still provide capacity gains. Increasing the antenna-movement DoFs of NC-MA, however, does not translate into capacity improvement. Instead, NC-MA with $d_{\min} = 0.2\lambda$ achieves a significantly lower capacity than NC-MA with $d_{\min} = \lambda/2$, because MC effects introduce a mismatch between the modeled capacity and the physically achievable capacity. We also observe that the capacity of C-MA with $d_{\min} = \lambda/2$ is nearly identical to that of C-MA with $d_{\min} = 0.2\lambda$ when $p>0.5$. This result indicates that even with the same antenna-movement DoFs as NC-MA, C-MA can still attain capacity gains through superdirectivity and adaptive MC matrix shaping.

\subsection{Wideband System}
\begin{figure}[!t]
    \centering
    \includegraphics[width=0.45\textwidth]{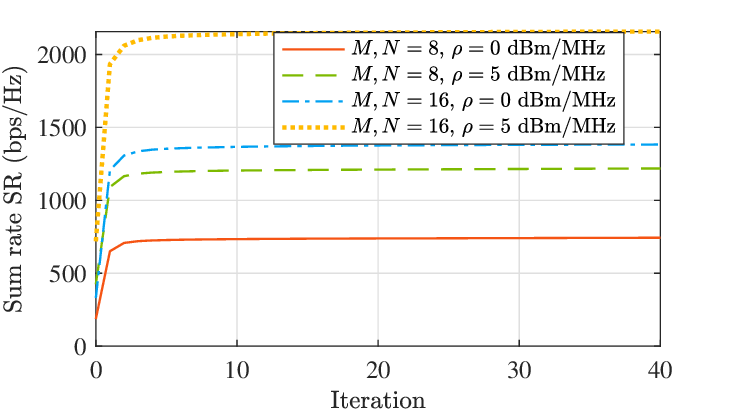}
    \caption{Convergence behaviors of Algorithm~\ref{alg:ofdm_bca} in wideband scenarios.}
    \label{fig:ofdm_conv}
\end{figure}

\begin{figure}[!t]
    \centering
    \includegraphics[width=0.45\textwidth]{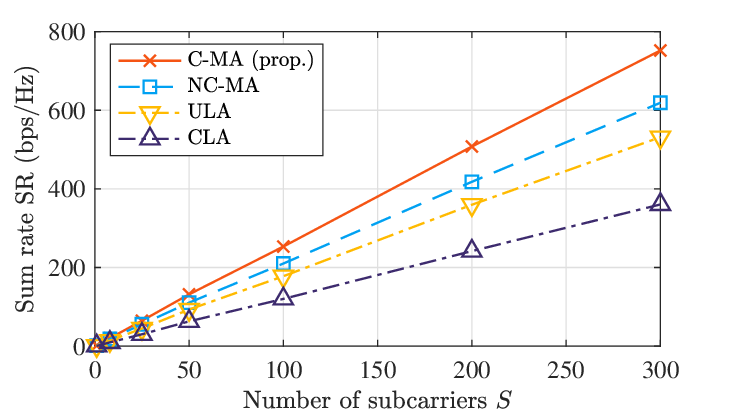}
    \caption{Impact of number of subcarriers $S$ on the sum-rate $\mathrm{SR}$ of wideband MC-aware MA systems.}
    \label{fig:ofdm_nsub}
\end{figure}

Fig.~\ref{fig:ofdm_conv} illustrates the convergence behavior of Algorithm~\ref{alg:ofdm_bca} in wideband scenarios. The sum-rate $\mathrm{SR}$ achieved by Algorithm~\ref{alg:ofdm_bca} increases monotonically with the number of iterations for different antenna numbers $M$ and $N$ and transmit power densities $\rho$. Algorithm~\ref{alg:ofdm_bca} converges within approximately $20$ iterations in all considered setups. The impact of the number of subcarriers $S$ on the sum-rate $\mathrm{SR}$ of MC-aware MA systems in wideband scenarios is illustrated in Fig.~\ref{fig:ofdm_nsub}. All schemes exhibit an approximately linear increase in sum-rate $\mathrm{SR}$ with the number of subcarriers $S$. The sum-rates of ULA and CLA scale linearly with $S$ because the transmit covariance matrix can be optimized independently on each subcarrier, thereby achieving almost invariant per-subcarrier capacities. The similar scaling behaviors of C-MA and NC-MA indicate that antenna movement in MA systems can effectively accommodate the varying channel conditions across subcarriers. Moreover, the sum-rate gain of C-MA over NC-MA becomes more pronounced as $S$ increases. This is because superdirectivity and designable MC matrices enhance capacity across subcarriers. 

\begin{figure}[!t]
    \centering
    \includegraphics[width=0.45\textwidth]{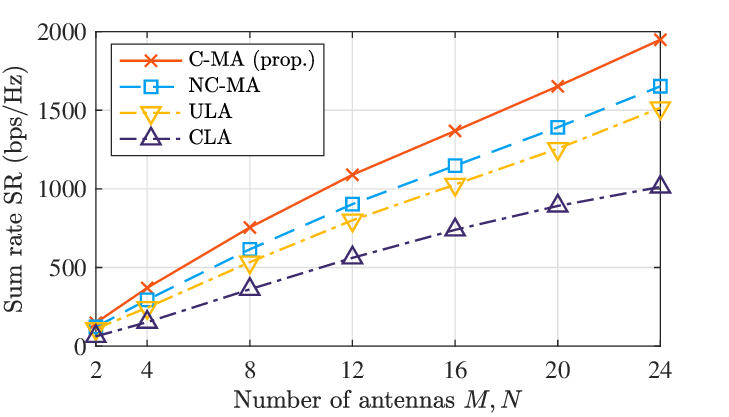}
    \caption{Impact of number of transmit MAs $M$ and receive MAs $N$ on the sum-rate $\mathrm{SR}$ of wideband MC-aware MA systems.}
    \label{fig:ofdm_ntx}
\end{figure}

\begin{figure}[!t]
    \centering
    \includegraphics[width=0.45\textwidth]{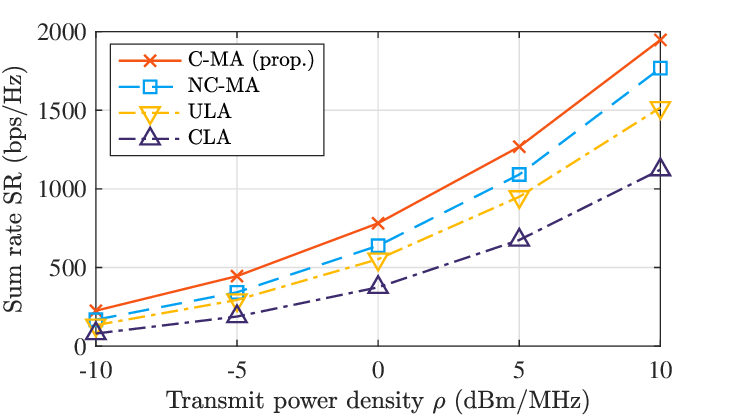}
    \caption{Impact of transmit power density $\rho$ on the sum-rate $\mathrm{SR}$ of wideband MC-aware MA systems.}
    \label{fig:ofdm_power}
\end{figure}

The impact of the numbers of transmit MAs $M$ and receive MAs $N$ on the sum-rate $\mathrm{SR}$ is depicted in Fig.~\ref{fig:ofdm_ntx}. When $M = N = 8$, the sum-rate of C-MA exceeds that of NC-MA by approximately $187\,$bps/Hz. When $M = N = 24$, this gap increases to approximately $296\,$bps/Hz. Fig.~\ref{fig:ofdm_power} illustrates the impact of the transmit power density $\rho$ on the sum-rate $\mathrm{SR}$. The variation of $\mathrm{SR}$ with $\rho$ follows a trend similar to that observed in the narrowband scenario shown in Fig.~\ref{fig:nb_power}. When $\rho = 0$\,dBm/MHz, the sum-rate of C-MA is approximately $22.4\,\%$ higher than that of NC-MA. When $\rho = 10$\,dBm/MHz, this gain decreases to about $10.2\,\%$. For different transmit power densities and numbers of antennas, the sum-rate achieved by C-MA is consistently higher than those of all baseline schemes. 

\subsection{How MC Improves Capacity in MA Systems}
As suggested by Remark~\ref{rem:capacity_gain}, the capacity gains of C-MA over NC-MA stem from superdirectivity and designable MC matrices. In this subsection, we verify these effects numerically using the parameters in Table~\ref{tab:sim_params}. Since the sum-rate in wideband scenarios is obtained by summing the capacities over all subcarriers, it suffices to examine the narrowband capacity gains of C-MA over NC-MA. Moreover, because the transmit and receive arrays are symmetric, we focus only on the superdirective beam gains at the transmitter. 

Compared with NC-MA, the capacity gains of C-MA originate from the superdirectivity enabled by MC. To verify this observation, we evaluate the transmitted power density along the directions $\{\theta_l^{\mathtt{T}}\}_{l=1}^{L}$:
\begin{equation*}
    P_{\mathtt{trans}} = \trace\left(\mathbf{G}(\mathbf{t}) \mathbf{C}_{\mathtt{T}}^{-1/2}(\mathbf{t}) \mathbf{Q} \mathbf{C}_{\mathtt{T}}^{-1/2}(\mathbf{t}) \mathbf{G}^\hermconj(\mathbf{t})\right).
\end{equation*}
For NC-MA and C-MA, the values of $P_{\mathtt{trans}}$ are $52.8$~W/rad and $64.9$~W/rad, respectively. The higher $P_{\mathtt{trans}}$ in C-MA over NC-MA leads to a higher capacity. However, superdirectivity alone cannot guarantee high capacity. Although CLA exhibits the strongest superdirectivity among all schemes, its capacity is the lowest. The strength of superdirectivity is characterized by the quality factor, defined as the reciprocal of the smallest eigenvalue of the MC matrix~\cite{morrisSuperdirectivityMIMOSystems2005}. The quality factors of CLA, C-MA, and NC-MA are $9.27\times 10^5$, $1.60$, and $1.16$, respectively. Despite its extremely strong superdirectivity, the small aperture of CLA severely limits its angular resolution, thereby preventing efficient power injection toward the desired directions. In other words, although the eigenvalues of $\mathbf{C}_{\mathtt{T}}^{-1/2}$ in CLA are significantly larger than those in C-MA, the corresponding eigenvectors are not well aligned with those of the FRM $\mathbf{G}$. Consequently, the value of $P_{\mathtt{trans}}$ in CLA is $54.0$~W/rad, which is lower than that in C-MA. This result indicates that attaining high capacity requires not only superdirectivity but also proper shaping of the MC matrices to better match the wireless channel.

\section{Conclusion}\label{sec:conclusion}
In this paper, we proposed to leverage MC in an MA-enabled point-to-point MIMO system. To capture the underlying electromagnetic effect, the system was characterized using a circuit-based communication model. Building upon this physically accurate model, we proposed a comprehensive optimization framework to unlock the potential of MC effects. First, we formulated the narrowband capacity maximization problem as a non-concave optimization problem and solved it using a BCA-based framework. For MA position optimization, we developed a TRM-based algorithm and derived the first- and second-order derivatives using Sylvester equations. We then considered a wideband communication system and formulated the wideband sum-rate maximization problem as a non-concave optimization problem. Next, we formulated a more challenging wideband joint covariance matrix and MA position optimization problem. This problem was solved by extending the BCA-based framework and the TRM-based algorithm. Simulation results demonstrated that exploiting MC in MA-MIMO systems achieves significant performance gains in both narrowband and wideband scenarios, even under strong LoS conditions and limited antenna movement ranges. These gains were shown to originate from superdirectivity and designable MC matrices. Promising future research directions include incorporating more realistic antenna models such as dipoles, developing low-complexity optimization algorithms, and investigating multiuser communication scenarios.

\bibliographystyle{IEEEtran}
\bibliography{IEEEabrv,reference}

\end{document}